\documentclass[a4paper,11pt]{article}
\usepackage{amssymb,float,amsmath,amsfonts,amsthm,cases,cite,bm,latexsym,mathrsfs}
\usepackage[normalem]{ulem}
\usepackage{graphicx,color,epsfig}
\usepackage{wrapfig}
\usepackage[pdfencoding=auto]{hyperref}
\usepackage[dvipsnames]{xcolor}
\usepackage{hyperref}
\definecolor{linkcolor}{HTML}{799B03}
\definecolor{urlcolor}{HTML}{799B03}
\hypersetup{pdfstartview=FitH,linkcolor=linkcolor,urlcolor=urlcolor,colorlinks=true}
\newcommand*{\D}{{\rm d}}

\def\[{\begin{equation}}
\def\]{\end{equation}}

\begin{document}	

\vspace{10pt}

\begin{center}
{\LARGE \bf More about stable wormholes\\ in beyond Horndeski theory}

\vspace{20pt}
S. Mironov$^{a,c}$\footnote{sa.mironov\_1@physics.msu.ru},
V. Rubakov$^{a,b}$\footnote{rubakov@inr.ac.ru},
V. Volkova$^{a,b}$\footnote{volkova.viktoriya@physics.msu.ru}
\renewcommand*{\thefootnote}{\arabic{footnote}}
\vspace{15pt}

$^a$\textit{Institute for Nuclear Research of the Russian Academy of Sciences,\\
60th October Anniversary Prospect, 7a, 117312 Moscow, Russia}\\
\vspace{5pt}

$^b$\textit{Department of Particle Physics and Cosmology, Physics Faculty,\\
M.V. Lomonosov Moscow State University,\\
Vorobjevy Gory, 119991 Moscow, Russia}

$^c$\textit{Institute for Theoretical and Experimental Physics,\\
Bolshaya Cheriomyshkinskaya, 25, 117218 Moscow, Russia}
\end{center}

\vspace{5pt}

\begin{abstract}

 {It is known that Horndeski theories, like many other scalar-tensor
  gravities,
  do not support static, spherically symmetric wormholes:
  they always have either ghosts or gradient instabilities among parity-even
  linearized perturbations.
  Here we address the issue of whether or not this no-go theorem is valid in 
  ``beyond Horndeski'' theories. We derive, in the latter class
  of theories, the conditions for the
  absence of ghost and gradient
 instabilities  for non-spherical 
  parity even perturbations propagating in radial direction.
  We find, in agreement with existing arguments,
  that the proof of the above no-go theorem does not go through
  beyond Horndeski.
  We also obtain  conditions ensuring the absence of ghosts and
  gradient instabilities   for all parity odd modes.
    We give an example of beyond Horndeski Lagrangian which admits
  a wormhole solution obeying our (incomplete set of) stability conditions.
Even though our stability analysis is incomplete, as we do not
consider spherically symmetric parity even modes and parity
even perturbations propagating in angular directions, as well as ``slow''
tachyonic instabilities, our findings indicate
that beyond Horndeski theories may be viable candidates to support
traversable wormholes.}
\end{abstract}

\section{Introduction}
\label{sec:intro}

Traversable wormholes are intriguing objects whose hypothetical
existence has fascinated the scientific community for many
years~\cite{Ellis, Bronnikov, Morris:1988cz, Morris:1988tu, Visser}.
Within General Relativity,
one of the
non-trivial features of the traversable wormholes is the necessity
to fill the space near the throat with  exotic matter, which
violates the Null Energy Condition (NEC).
Since the NEC is quite robust and the majority of forms of matter
comply with it, obtaining stable traversable wormhole 
is challenging.
{As an example, wormhole solutions exist in
theories involving scalar field with wrong sign kinetic
term~\cite{Ellis, Bronnikov,  Morris:1988cz, Morris:1988tu, Visser,Kodama, Armendariz}, but these ghost
theories are catastrophically unstable in quantum
theory~\cite{Cline:2003gs} (see Ref.~\cite{Rubakov:2014jja} for a review).
Once gravity is modified, the NEC is no longer relevant; one replaces
it with the Null Convergence Condition (NCC)~\cite{Tipler, Steinhardt},
which says that
$R_{\mu \nu} n^\mu n^\nu > 0$
for all null vectors $n^\mu$. The NCC is also hard to
violate~\footnote{ There is a claim~\cite{Kanti1, Kanti2} that there exists
  a stable static wormhole in a dilaton--Gauss--Bonnet
  gravity. This solution, however, appears unstable even against
  spherically symmetric perturbations~\cite{Cuyubamba:2018jdl}. Furthermore,
  the construction  of Refs. ~\cite{Kanti1, Kanti2}
  involves a  negative tension
  2-brane 
  at the wormhole throat. It is known in other contexts that a brane bending
  mode  (not considered in Refs. ~\cite{Kanti1, Kanti2,Cuyubamba:2018jdl})
  in the perturbation spectrum of a negative tension
  brane 
  tends to be a  ghost (see, e.g.,
  Ref.~\cite{Pilo}). Also, the stability analysis of the
non-spherical perturbations about the
  wormhole of Refs.~\cite{Kanti1, Kanti2} is still lacking.}. }

Recently, it has been suggested that it might be possible to construct
a
wormhole by employing Horndeski theories~\cite{Horndeski:1974wa},
which are remarkable due to their ability to violate the NEC and NCC
without
introducing ghost or gradient instabilities.
Horndeski theory is a scalar-tensor gravity
whose Lagrangian involves terms with second derivatives but the
equations of motion remain second order. There is
an extension of Horndeski theories usually referred to as ``beyond
Horndeski''
theories~\cite{Gleyzes:2014dya}.
The major difference between the general Horndeski theory and
its extension is that the latter has  equations of motion of the
third order in derivatives. Nevertheless,
the
number of dynamical degrees of freedom is the same in Horndeski and
beyond Horndeski theory ~\cite{Gleyzes:2014qga}.

Since (beyond) Horndeski theories are capable of violating the NEC
{and NCC without
introducing ghost or gradient instabilities},
they have become quite popular in studies
of  scenarios requiring the NEC{/NCC}
violation, e.g., the Universe with a
bounce{, see Ref.~\cite{Kobayashi-2019} for review}.
Attempts to construct  traversable wormholes within
various subclasses of Horndeski theories were made, for instance,
in Refs.~\cite{Bronnikov:2010tt, Korolev:2014hwa, Rubakov:2015gza, Rubakov:2016zah, Kolevatov:2016ppi}, but stability of the solutions
has always been a troublesome issue. Finally,
a no-go theorem has been proven in  Ref.~\cite{Olegi},
which forbids the existence of a
stable, static, spherically symmetric wormhole solution in the general
Horndeski theory. Namely, 
static, spherically symmetric wormholes
in Horndeski theory
inevitably have a region (typically near the throat)
where non-spherical
parity even (symmetric under reflection of 2-sphere)
modes of linear perturbations
have wrong sign time-derivative term in their quadratic action,
meaning either
ghost or gradient instability. Note that this instability is local
in space, as it is associated with high spatial momenta,
and that it is catastrophic, since
frequencies of unstable modes are arbitrarily high.

Interestingly, this no-go theorem has its close analogue
in the cosmological setting, namely, similar no-go argument holds
for non-singular homogeneous, isotropic, spatially flat
solutions in Horndeski theory,
for instance, the cosmological bounce~\cite{Libanov:2016kfc,Kobayashi:2016xpl}.
The analogy between the cosmological bounce and wormhole
is not utterly surprising: the radial profile of a wormhole
resembles the evolution of the scale factor in the bouncing Universe.
Although imperfect, this wormhole/bounce analogy becomes even more
tempting because of the recent resolution of the stability issue in
non-singular cosmological scenarios: in that case,
the no-go theorem has been evaded~\cite{Cai:2016thi,Creminelli:2016zwa}
by going beyond Horndeski, and complete stable bouncing solutions have
been constructed~\cite{Kolevatov:2017voe,Cai:2017dyi,Mironov:2018oec}.
One may conjecture {that beyond Horndeski theories may admit stable
wormholes 
as well. A necessary (possibly, crucial) step
in checking this conjecture
is to construct a wormhole in a beyond Horndeski theory
that explicitly evades the no-go theorem proven in the Horndeski case.}
Such a construction has been attempted in Ref.~\cite{Mironov:2018pjk}
where we presented static, spherically symmetric
wormhole in beyond Horndeski
theory which, we argued, {did not have
ghost instabilities
among parity even modes, and also among
parity odd modes}\footnote{Ref.~\cite{Mironov:2018pjk}
  had a computational error
  which lead to a wrong conclusion concerning fine-tuning. We correct the
  error in this paper.}. Independently, by making use of the
{effective field theory (EFT)
approach
based on ADM formalism},
it was shown in
Ref.~\cite{Trincherini}
that
the proof of the no-go theorem {does not go through
in beyond Horndeski theory, and that the parity odd modes may be
completely stable.}

In this paper, we obtain part of the stability conditions for a static,
spherically symmetric solution in beyond Horndeski theory in a
covariant form (as opposed to ADM EFT form of Ref.~\cite{Trincherini}).
Namely, we derive  the conditions ensuring
the absence of ghosts and absence of gradient
instabilities for non-spherical parity even modes propagating in radial
direction. We also find the conditions for the absence of
ghosts and gradient instabilities for all
parity odd modes.
These results extend the stability analysis
carried out for Horndeski theory
in Refs.~\cite{Kobayashi:odd, Kobayashi:even} to the beyond Horndeski
case; note that neither we nor
Refs.~\cite{Trincherini,Kobayashi:odd, Kobayashi:even} study
the full set of stability issues
which includes the absence of gradient instability in angular directions
in parity even {sector as well as
the absence of ``slow'' tachyonic modes.}
Yet our analysis is sufficient to
show explicitly that beyond Horndeski terms modify the stability
conditions in such a way that the no-go argument of Ref.~\cite{Olegi} no longer
applies. Not surprizingly, the form of the modified stability
conditions in spherically symmetric background is similar to that in
the homogeneous cosmological setting,
which deepens the analogy between the two cases.

We then suggest a model  with a specific Lagrangian of beyond
Horndeski type, which admits a wormhole solution
free of all {those}
instabilities that we consider in this paper.
This explicit example (which, in fact, is {inspired by}
our original model given in Ref.~\cite{Mironov:2018pjk})
shows that the no-go theorem can indeed be
circumvented in the beyond Horndeski theories. However,
the example is not
completely satisfactory: even though space-time is {asymptotically
Minkowskian,
weak gravity regime is grossly different from
General Relativity
even at large distances from the wormhole.}
Yet another possibly
unphysical feature is that our wormhole has vanishing mass.

{Thus, we give both general arguments and explicit example showing
that the obstacle for constructing
stable, static, spherically symmetric wormholes in Horndeski theories is
absent in beyond Horndeski case. We emphasize, however, that we do not
pretend to construct a fully stable wormhole. Indeed, we do not
consider spherically symmetric parity even modes and parity
even perturbations propagating in angular directions, as well as ``slow''
tachyonic instabilities. Note that the latter
instabilities
are potentially quite dangerous: as an example, they
ruin classical linear stability of
wormholes in theories with ghost fields~\cite{Gonzalez:2008wd}
(over and beyond quantum catastrophic instability of these theories),
whose geometries are very similar to that of our wormhole.
Yet we think that our findings indicate
that beyond Horndeski theories may be viable candidates to support
traversable wormholes.}

This paper is organized as follows. We give generalities of the
beyond Horndeski theory and spherically symmetric Ansatz in
section~\ref{set_up}, derive our {subset of}
stability conditions and show
that the wormhole no-go theorem can be circumvented in
section~\ref{linearized_theory}. {We}
construct the wormhole solution
obeying our subset of stability conditions in section~\ref{sec:wormhole}.
We conclude in section~\ref{sec:conclusion}.
Explicit formulas, which are often cumbersome, are collected in
Appendices.

\section{Beyond Horndeski Lagrangian}
\label{set_up}
In what follows we generalize the stability analysis of
Refs.~\cite{Kobayashi:odd,Kobayashi:even} by introducing the
beyond Horndeski terms in the Lagrangian. The most general form of
the beyond Horndeski Lagrangian is (with mostly positive
signature of the metric)
\begin{subequations}
	\label{lagrangian}
	\begin{align}
	&\phantom{\mathcal{L}_2=F(\pi,X)}S=\int\mathrm{d}^4x\sqrt{-g}\left(\mathcal{L}_2 + \mathcal{L}_3 + \mathcal{L}_4 + \mathcal{L}_5 + \mathcal{L_{BH}}\right),\\
	&\mathcal{L}_2=F(\pi,X),\\
	&\mathcal{L}_3= - K(\pi,X)\Box\pi,\\
	&\mathcal{L}_4=G_4(\pi,X)R + G_{4X}(\pi,X)\left[\left(\Box\pi\right)^2-\pi_{;\mu\nu}\pi^{;\mu\nu}\right],\\
	\label{L5}
	&\mathcal{L}_5= G_5(\pi,X)G^{\mu\nu}\pi_{;\mu\nu}-\frac{1}{6}G_{5X}\left[\left(\Box\pi\right)^3-3\Box\pi\pi_{;\mu\nu}\pi^{;\mu\nu}+2\pi_{;\mu\nu}\pi^{;\mu\rho}\pi_{;\rho}^{\;\;\nu}\right],\\
	\label{BH}
	&\mathcal{L_{BH}}=F_4(\pi,X)\epsilon^{\mu\nu\rho}_{\quad\;\sigma}\epsilon^{\mu'\nu'\rho'\sigma}\pi_{,\mu}\pi_{,\mu'}\pi_{;\nu\nu'}\pi_{;\rho\rho'}+
	\\\nonumber&\qquad+F_5(\pi,X)\epsilon^{\mu\nu\rho\sigma}\epsilon^{\mu'\nu'\rho'\sigma'}\pi_{,\mu}\pi_{,\mu'}\pi_{;\nu\nu'}\pi_{;\rho\rho'}\pi_{;\sigma\sigma'},
	\end{align}
\end{subequations}
where $\pi$ is a scalar field (sometimes dubbed generalized Galileon),
$X= -\frac12\: g^{\mu\nu}\pi_{,\mu}\pi_{,\nu}$,
$\pi_{,\mu}=\partial_\mu\pi$,
$\pi_{;\mu\nu}=\triangledown_\nu\triangledown_\mu\pi$,
$\Box\pi = g^{\mu\nu}\triangledown_\nu\triangledown_\mu\pi$,
$G_{iX}=\partial G_i/\partial X$.
Functions $F_4(\pi,X)$ and $F_5(\pi,X)$ in~\eqref{BH} are
characteristic of beyond Horndeski case, while the general Horndeski
theory involves $\mathcal{L}_2 - \mathcal{L}_5$.
In this paper we set \[
F_5(\pi,X)=0 \; ,
\]
since the
function $F_4(\pi,X)$
is sufficient to reveal the difference between Horndeski and
beyond Horndeski cases.
When generalizing the stability conditions found in
Refs.~\cite{Kobayashi:odd, Kobayashi:even} to beyond Horndeski
theory, we intensely use their notations. In what follows we set
\[
8\pi M_{Pl}^2 =1 \; .
\]

We consider static, spherically symmetric background geometry
with  metric of the following form:
\[
\label{backgr_metric}
ds^2 = - A(r)\:dt^2 + \frac{dr^2}{B(r)} + J^2(r) \left(d\theta^2 + \sin^2\theta\: d\varphi^2\right),
\]
where the radial coordinate $r$ runs from $-\infty$ to $+\infty$,
and the functions $A(r)$, $B(r)$ and $J(r)$ are positive and bounded
from below,
\[
A(r) \geq A_{min} > 0, \quad B(r) \geq B_{min} > 0, \quad J(r) \geq R_{min} > 0,
\]
with $R_{min}$ standing for the radius of the wormhole throat.
{Note that $r$ is a globally defined coordinate, so that the
horizon absence and  ``flaring-out''
conditions are satisfied automatically.
Wormhole is asymptotically flat at both sides provided that
\[
A(r) \to 1 \; , \;\;\;\; B(r) \to 1 \; ,
\;\;\;\; J(r) = |r| + O(1) \;\;\;\; \mbox{as} \;\;\; r \to \pm \infty \; .
\]
}
Even though $B(r)$  in eq.~\eqref{backgr_metric}
can be set equal to 1 by coordinate transformation,
we keep it arbitrary.
The scalar field $\pi$ is  static and
depends on the radial coordinate only, $\pi = \pi(r)$ and
hence $X = -B(r) \pi'^2 /2$, where prime denotes the
derivative with respect to $r$.

To carry out the stability analysis,
we adopt a bottom up approach, namely, we derive the background
equations for the action~\eqref{lagrangian} first, linearize them
and then reconstruct the quadratic action for perturbations.
We keep the notations of Ref.~\cite{Kobayashi:even} for the
equations of motion:
\[
  {\cal E}_A = 0, \quad {\cal E}_B = 0, \quad {\cal E}_J = 0,
  \quad {\cal E}_{\pi} = 0 \; ,
  \]
  where $ {\cal E}_A$ is obtained by varying the action with respect to
  $A$, etc.
The explicit forms of
${\cal E}_A, {\cal E}_B, {\cal E}_J$ and ${\cal E}_{\pi}$
are given
in Appendix A, eqs.~\eqref{CalEa}--\eqref{CalEphi}.
In the next section we introduce the parametrization of
perturbations and develop the corresponding linearized theory.

\section{Linearized theory}
\label{linearized_theory}
\subsection{Parametrizing the perturbations}
\label{notations}
In the same manner as in
Refs.~\cite{Kobayashi:odd,Kobayashi:even}, we make use of
the Regge--Wheeler classification of perturbations and decompose them
into parity
odd and parity even sectors~\cite{Regge}, according to their transformation
laws
under the two-dimensional reflection. Perturbations are
further expanded in spherical
harmonics $Y_{\ell m}(\theta,\varphi)$.
This approach to parametrization simplifies calculations, since not
only the odd and even modes evolve independently, but also perturbations
with different $\ell$ and $m$ do not mix at the linearized level.

The perturbed metric reads
\[
g_{\mu\nu} = \bar{g}_{\mu\nu} + h_{\mu\nu},
\]
where $\bar{g}_{\mu\nu}$ stands for the background
metric~\eqref{backgr_metric} and $h_{\mu\nu}$ are linear perturbations.
The parity odd sector of metric perturbations has the following
parametrization:
\[
\label{odd_parity}
\mbox{Parity~odd}\quad \begin{cases}
\begin{aligned}
 & h_{tt}=0,~~~h_{tr}=0,~~~h_{rr}=0,\\
 & h_{ta}=\sum_{\ell, m}h_{0,\ell m}(t,r)E_{ab}\partial^{b}Y_{\ell m}(\theta,\varphi),\\
 & h_{ra}=\sum_{\ell, m}h_{1,\ell m}(t,r)E_{ab}\partial^{b}Y_{\ell m}(\theta,\varphi),\\
 & h_{ab}=\frac{1}{2}\sum_{\ell, m}h_{2,\ell m}(t,r)\left[E_{a}^{~c}\nabla_{c}\nabla_{b}Y_{\ell m}(\theta,\varphi)+E_{b}^{~c}\nabla_{c}\nabla_{a}Y_{\ell m}(\theta,\varphi)\right],
\end{aligned}
\end{cases}
\]
where $a,b = \theta,\varphi$, $E_{ab} = \sqrt{\det \gamma}\: \epsilon_{ab}$, with $\gamma_{ab} = \mbox{diag}(1, \;\sin^2\theta)$;
$\epsilon_{ab}$ is
totally antisymmetric symbol ($\epsilon_{\theta\varphi} = 1$) and $\nabla_a$
is covariant derivative on 2-sphere.
Decomposition
in the parity even sector reads
\[
\label{even_parity}
\mbox{Parity~even}\quad \begin{cases}
\begin{aligned}
h_{tt}=&A(r)\sum_{\ell, m}H_{0,\ell m}(t,r)Y_{\ell m}(\theta,\varphi), \\
h_{tr}=&\sum_{\ell, m}H_{1,\ell m}(t,r)Y_{\ell m}(\theta,\varphi),\\
h_{rr}=&\frac{1}{B(r)}\sum_{\ell, m}H_{2,\ell m}(t,r)Y_{\ell m}(\theta,\varphi),\\
h_{ta}=&\sum_{\ell, m}\beta_{\ell m}(t,r)\partial_{a}Y_{\ell m}(\theta,\varphi), \\
h_{ra}=&\sum_{\ell, m}\alpha_{\ell m}(t,r)\partial_{a}Y_{\ell m}(\theta,\varphi), \\
h_{ab}=&\sum_{\ell, m} K_{\ell m}(t,r) g_{ab} Y_{\ell m}(\theta,\varphi)+\sum_{\ell, m} G_{\ell m}(t,r) \nabla_a \nabla_b Y_{\ell m}(\theta,\varphi)\,.
\end{aligned}
\end{cases}
\]
The perturbation of the scalar field $\pi$ is non-vanishing
only in the parity even sector:
\[
\label{chi}
\pi (t,r,\theta,\varphi) = \pi(r) + \sum_{\ell, m}\chi_{\ell m}(t,r)Y_{\ell m}(\theta,\varphi),
\]
where $\pi(r)$ is the spherically symmetric background field.

{To proceed further, one either makes use of gauge-invariant
variables~\cite{Gerlach:1979rw} or imposes gauge conditions.
We follow the latter route and}
adopt
the gauge choice made in Refs.~\cite{Kobayashi:odd,Kobayashi:even},
namely, in the parity odd sector we set
\[
h_{2,\ell m}(t,r) =0\; ,
\label{dec10-18-2}
\]
while in
the parity even sector we impose the gauge conditions
\[
\beta_{\ell m}(t,r)=0\; , \;\;\;\;\; K_{\ell m}(t,r)=0
\; , \;\;\;\;\;
G_{\ell m}(t,r)= 0 \; .
\label{dec10-18-1}
\]
Note that our gauge
choice differs from that of Ref.~\cite{Trincherini},
but despite this fact we expect to have the same linearized
spectrum in the end.

We consider the linearized theory for parity odd and parity even
perturbations separately 
in the next two  subsections.

\subsection{Parity odd sector}
\label{quad_action_odd}
{To warm up, we consider parity odd sector first.}
As outlined above, to obtain the quadratic action for
perturbations we linearize equations of motion
(eqs.~\eqref{CalEa}--\eqref{CalEphi} in Appendix A) and
reconstruct the corresponding quadratic action.
{Note that the perturbations with $\ell=0$ do not exist, while
perturbations with $\ell=1$ are pure gauges~\cite{Armendariz,Kobayashi:odd}.
So, we are left with
\[
\ell>1 \; .
\]
}
Making use of the notations \eqref{odd_parity}
for  parity odd
perturbations and gauge condition
\eqref{dec10-18-2}
and performing integration
by parts whenever necessary, we obtain the quadratic
action for the parity odd sector:
\[
\label{action_odd}
S^{(2)}_{odd} = \int \mbox{d}t\:\mbox{d}r \left[ A_1 h_0^2+A_2 h_1^2+A_3 \left( {\dot h_1}^2-2 {\dot h_1} h_0'+h_0'^2+\frac{4 J'}{J} {\dot h_1} h_0 \right) \right],
\]
where dot and prime stand for the temporal and radial
derivatives, respectively. Since modes with different angular
momenta
$\ell$ and $m$ decouple, hereafter we drop
subscripts $\ell$ and $m$ in
$h_0$,  $h_1$ and other perturbations; summation over $\ell$ and $m$
is implicit in \eqref{action_odd} and in similar expressions
below.
We integrated over the angular coordinates in
eq.~\eqref{action_odd}. The coefficients $A_1$, $A_2$ and
$A_3$ include $\sqrt{-g}$ and read
\begin{eqnarray}
\label{A1}
A_1&=&\frac{\ell (\ell+1)}{J^2}
\left[\frac{d}{d r}\left(
J\: J' \sqrt{\frac{B}{A}}{\cal H}\right)
+\frac{\ell^2+\ell-2}{2\sqrt{AB}}{\cal F}
+\frac{J^2}{\sqrt{AB}} {\cal E}_A \right],
\\
\label{A2}
A_2&=&- \ell (\ell+1) \sqrt{AB} \left[ \frac{(\ell-1) (\ell+2)}{2J^2}{\cal G}+{\cal E}_B \right],
\\
\label{A3}
A_3&=&
\frac{\ell (\ell+1)}{2} \sqrt{\frac{B}{A}} {\cal H}.
\end{eqnarray}
We retain the left hand sides of equations of motion
${\cal E}_A$ and ${\cal E}_B$ in the expressions
for $A_1$ and $A_2$ for book keeping purposes.
The coefficients
in eqs.~\eqref{A1}--\eqref{A3}
get modified
as compared to their counterparts in
Ref.~\cite{Kobayashi:odd}:
\begin{eqnarray}
{\cal F}&=&
2 \left ( G_4 + \frac12 B \pi' X' G_{5X} - X G_{5\pi} \right)
,
\label{cal_f}
\\
{\cal G}&=& 2 \left[G_4 - 2X G_{4X} +X\left(\frac{A'}{2 A} B \pi' G_{5X} +G_{5\pi}\right)
+ 4 X^2 F_4 \right]
,
 \label{cal_g}
\\
{\cal H}&=&2 \left[G_4 - 2X G_{4X} +X\left(\frac{B J' \pi'}{J} G_{5X} +G_{5\pi}\right)
+ 4 X^2F_4 \right] .\label{cal_h}
\end{eqnarray}
Both ${\cal H}$ and ${\cal G}$ involve the beyond Horndeski
function $F_4(\pi, X)$.

The expression~\eqref{action_odd} for the action shows that
$h_0$ is a
non-dynamical degree of freedom, but the corresponding
constraint is a second-order differential equation.
To avoid solving the differential equation as it is, it was suggested
in Ref.~\cite{DeFelice:2011ka} to make use of the following trick for
rewriting the action~\eqref{action_odd}
in terms of a single  variable. First, we gather the
terms in the action~\eqref{action_odd}
containing derivatives of $h_0$ and $h_1$,
 and introduce the Lagrange multiplier $Q$:
\[
\label{action_odd_Q}
S^{(2)}_{odd} = \int \mbox{d}t\:\mbox{d}r \left[ A_1 h_0^2+A_2 h_1^2 + A_3 \left( 2 Q \left( {\dot h_1}- h_0' +\frac{2 J'}{J} h_0\right) - Q^2 \right) - \frac{2}{J^2}\frac{d}{dr}\left(A_3 J' J \right) h_0^2 \right].
\]
The equations of motion for $h_0$ and $h_1$ following from eq.~\eqref{action_odd_Q} read, respectively:
\[
\label{h0_h1}
h_0 = \frac{d}{dr}\left( A_3 Q J^2 \right)\cdot \left[ 2 \frac{d}{dr}\left( A_3 J' J \right)- A_1 J^2\right]^{-1},\quad\quad
h_1 = \frac{A_3}{A_2} \dot{Q}.
\]
Upon substituting eqs.~\eqref{h0_h1} into the
action~\eqref{action_odd_Q} and making use of expressions
for coefficients
$A_i$ given by eqs.~\eqref{A1}--~\eqref{A3}, we obtain the
quadratic action in terms of the dynamical degree of freedom $Q$:
\[
\label{action_odd_final}
\begin{aligned}
S^{(2)}_{odd} = \int \mbox{d}t\:\mbox{d}r\:\sqrt{\frac{A}{B}} J^2
\frac{\ell(\ell+1)}{2(\ell-1)(\ell+2)}\cdot \frac{B}{A}\left[ \frac{\mathcal{H}^2}{A \mathcal{G}} \dot{Q}^2 - \frac{B \mathcal{H}^2}{\mathcal{F}} (Q')^2 -\frac{l(l+1)}{J^2}\cdot \mathcal{H} Q^2 - V(r) Q^2 \right] \; ,
\end{aligned}
\]
which is the desired result.
The original variables $h_0$ and $h_1$ are
found from
\eqref{h0_h1} as soon as $Q$ is known.
Note that the third term in eq.~\eqref{action_odd_final} corresponds
to the angular part of the Laplace operator and governs the stability
in the angular  direction, while $V(r)$ stands for the
''potential'' and reads:
\begin{equation}
\label{Vr}
\begin{aligned}
V(r) &=  \frac{B {\cal H}^2 }{{2\cal F}}
\left[\frac{{\cal F}'}{{\cal F}} \left(2\frac{{\cal H}'}{{\cal H}} -\frac{A'}{A}+\frac{B'}{B}+ 4\frac{J'}{J}\right) -  \frac{{\cal H}'}{{\cal H}} \left( -\frac{A'}{A}+3 \frac{B'}{B}+4 \frac{J'}{J}\right)\right.\\
&\left. - \left(\frac{A'^2}{A^2} -\frac{A'}{A} \frac{B'}{B} - \frac{A''}{A} +\frac{B''}{B} +4 \frac{B'}{B} \frac{J'}{J}-4 \frac{J'^2}{J^2}+ 4 \frac{J''}{J} \right)  - \frac{4}{J^2 B}\cdot \frac{\cal F}{\cal H} \right].
\end{aligned}
\end{equation}

{The conditions for the absence
of ghosts and gradient instabilities
in the parity odd sector} immediately
follow from the quadratic action~\eqref{action_odd_final}:
\begin{eqnarray}
\label{stability_G}
\mbox{No ghosts:} &\mathcal{G} > 0, \\
\label{stability_F}
\mbox{No radial gradient instabilities:} &\mathcal{F} > 0, \\
\label{stability_H}
\mbox{No angular gradient instabilities:} &\mathcal{H} > 0,
\end{eqnarray}
The sound speeds squared for the modes propagating in the
radial and angular directions are, respectively:
\[
\label{speed_odd}
c_r^2 = \frac{\mathcal{G}}{\mathcal{F}}, \quad c_{\theta}^2 = \frac{\mathcal{G}}{\mathcal{H}}.
\]
To ensure that all modes propagate at subluminal speed, one
requires that
\[
\mathcal{F} \ge \mathcal{G} > 0
\label{sublu-odd-1}
\]
and
\[
\mathcal{H} \ge \mathcal{G} > 0 \; .
\label{sublu-odd-2}
\]
These inequalities
are similar to their counterparts in
 the general Horndeski case. However, the expressions for
 ${\cal G}$ and ${\cal H}$ now involve the
beyond Horndeski contributions.

{We note that the conditions \eqref{stability_G} -- \eqref{stability_H}
do not guarantee that the parity odd sector is completely stable.
Indeed, there may occur ``slow'' tachyonic instabilities
due to possible negative eigenvalues of the pertinent eigenvalue problem
involving the ``potential'' $V$. It has been argued, though, that
requirement of the absence of these instabilities
may not be particularly restrictive~\cite{Trincherini}.}

\subsection{Parity even sector: circumventing the no-go
theorem}
\label{quad_action_even}
Let us now consider the parity even sector of perturbations.
{For technical reasons, we consider non-spherical perturbations,
$\ell \neq 0$; we do not expect anything particularly pathological
in $\ell=0$
sector\footnote{The case $\ell =1$ is special too~\cite{Kobayashi:even}.
  In that case, there is one dynamical mode, rather than two at $\ell >1$.
  Upon imposing appropriate gauge conditions, we find that
  the coefficients in its quadratic action similar to
  \eqref{even_action_final} are given by \eqref{K11} and \eqref{G11}.
  Therefore, our stability analysis is valid for $\ell=1$ as well.}.}
In full analogy with the parity odd case, we linearize the
equations of motion~\eqref{CalEa}--\eqref{CalEphi},
adopting the parametrization of the
perturbations~\eqref{even_parity} and~\eqref{chi} and gauge
conditions~\eqref{dec10-18-1}, and arrive at
the following quadratic action:
\begin{equation}
\label{action_even}
\begin{aligned}
&S_{even}^{(2)} = \int \mbox{d}t\:\mbox{d}r \left(H_0 \left[ a_1 \chi''+ a_2
\chi'+a_3 H_2'+j^2 a_4 \alpha'+\left( a_5+j^2 a_6 \right) \chi\right.\right.  \\
&\left.+\left( a_7+j^2 a_8 \right)H_2+j^2a_9 \alpha \right]
+j^2 b_1 H_1^2+H_1 \left[b_2 {\dot {\chi}}'+b_3 {\dot {\chi}}+b_4 {\dot H_2}+j^2b_5 {\dot \alpha}\right] \\
&+ c_1{\dot H_2} {\dot {\chi}} + H_2 \left[c_2  \chi'+\left( c_3+j^2 c_4\right)\chi + j^2 c_5 \alpha \right] + c_6 H_2^2+j^2d_1 {\dot \alpha}^2
\\
&\left.+j^2 d_2 \alpha\chi'+j^2 d_3 \alpha\chi+j^2 d_4 \alpha^2
+ e_1{\dot {\chi}}^2+e_2 \chi'^2+\left( e_3+j^2 e_4  \right) \chi^2\right),
\end{aligned}
\end{equation}
where the subscripts $\ell$, $m$ are again omitted, $j^2=\ell(\ell+1)$, and
we have integrated over $\theta$ and $\phi$. The
explicit expressions for coefficients $a_i$, $b_i$, $c_i$, $d_i$
and $e_i$ with $\sqrt{-g}$ included are given in Appendix B.
Unlike in the homogeneous case, there are no new structures
in the quadratic action arising due to adding the beyond Horndeski term.
But the expressions for the coefficients $a_2$, $a_6$, $c_1$,
$c_4$, $e_1$ and $e_4$ in~\eqref{action_even} change
significantly  (see Appendix B for details).
This substantially modifies the
stability conditions, as we show below.

According to the form of the action~\eqref{action_even},
$H_0$ is a Lagrange multiplier that gives the following constraint:
\[
\begin{aligned}
\label{H0}
a_1 \chi''+a_2  \chi'+a_3 H_2'+
j^2 a_4 \alpha'+\left( a_5+j^2 a_6 \right) \chi
+\left( a_7+j^2 a_8 \right)H_2+j^2a_9 \alpha = 0.
\end{aligned}
\]
Following Ref.~\cite{Kobayashi:even} we introduce a new
variable $\psi$ such that
\begin{equation}
\label{H2}
H_2 = \psi - \frac{1}{a_3}\left( a_1 \chi' + j^2 a_4 \alpha\right),
\end{equation}
then both $\chi''$ and $\alpha'$ terms get cancelled out in
eq.~\eqref{H0} upon substitution of $H_2$ from eq.~\eqref{H2}.
The resulting equation can be solved to give\footnote{It appears that
  there is a misprint in eq.(29)
  of Ref.~\cite{Kobayashi:even}.}
\begin{equation}
\label{alpha}
\alpha = \frac{a_3^2 \psi' + a_3(a_7+j^2a_8) \psi + [a_3 (a_2 - a_1') -j^2 a_1 a_8] \chi' + a_3 (a_5 + j^2 a_6) \chi}{j^2 \left[ a_3 a_4' - a_3' a_4 - a_3 a_9 + a_4 (a_7 + j^2 a_8\right]} \; ,
\end{equation}
where we made use of the relation $a_3' = a_7$ (see Appendix B).
$H_1$ is a also a non-dynamical degree of freedom in the
action~\eqref{action_even}. The corresponding constraint
 reads
\[
\label{H1}
 H_1 = - \frac{1}{2 j^2 b_1} \left( b_2 {\dot {\chi}}'+b_3 {\dot {\chi}}+b_4 {\dot H_2}+j^2b_5 {\dot \alpha}\right).
\]
Equation~\eqref{alpha} enables one to express $H_2$ and $H_1$,
using eqs.~\eqref{H2} and~\eqref{H1}, in terms of
$\psi$ and $\chi$.
Hence, upon substituting eqs.~\eqref{H2}-\eqref{H1} into
the action~\eqref{action_even} and integrating by parts,
the quadratic action is written  in terms
of two dynamical degrees of freedom:
\[
\label{even_action_final}
S_{even}^{(2)} = \int \mbox{d}t\:\mbox{d}r \sqrt{\frac{A}{B}} J^2 \left(
\frac12 \mathcal{K}_{ij} \dot{v}^i \dot{v}^j - \frac12 \mathcal{G}_{ij} v^{i\prime} v^{j\prime} - \mathcal{Q}_{ij} v^i v^{j\prime} - \frac12 \mathcal{M}_{ij} v^i v^j \right),
\]
where $i=1,2$ and $v^1=\psi$, $v^2=\chi$. We note that
terms which are higher order in derivatives, like
$\dot{\psi}' \dot{\chi}$,
safely disappear upon integrating by parts.

To ensure the absence of ghosts in the parity
even sector, we require that the quadratic form ${\cal K}_{ij}$
is positive definite, i.e.,
\[
\label{no_ghost}
\mathcal{K}_{11}>0, \qquad \det(\mathcal{K}) > 0.
\]
The explicit expressions
for $\mathcal{K}_{11}$ and $\det(\mathcal{K})$ read
\begin{equation}
\label{K11}
{\cal K}_{11}=\frac{8  B {\left( 2{\cal H} J J' +
\Xi \pi' \right)}^2 \left[ \ell (\ell+1){\cal P}_1-{\cal F}\right]}
{\ell (\ell+1) A^2 \mathcal{H}^2  \left[2 J \mathcal{H} \ell (\ell+1) +\mathcal{P}_2 -4 J F_4 \ell (\ell+1)  B^2 \pi'^4\right]^2},
\end{equation}
\begin{equation}
\label{detK}
\det ({\cal K})=\frac{16 B J'^2 (\ell-1) (\ell+2){\left( 2{\cal H} J J' + \Xi \pi'\right)}^2 \left[{\cal F}(2{\cal P}_1-{\cal F})\right]}
{\ell (\ell+1) A^3 \mathcal{H}^2 J^2 \pi'^2 \left[2 J \mathcal{H} \ell (\ell+1) +\mathcal{P}_2 -4 J F_4 \ell (\ell+1)  B^2 \pi'^4\right]^2},
\end{equation}
where ${\cal F}$ and $\mathcal{H}$ are given by \eqref{cal_f} and \eqref{cal_h},
\[
	\begin{aligned}
\label{Xi}
\Xi &= K_{X}BJ^2\pi'^2 + 2G_{4\pi}J^2 + 4G_{4X}BJJ'\pi'
 -2G_{4\pi X}BJ^2\pi'^2
  \\&
 - 4G_{4XX}B^2JJ'\pi'^3 -
    4G_{5\pi}BJJ'\pi' + G_{5X}B\pi'^2
    - 3G_{5X}B^2J'^2\pi'^2
  \\&
    + 2G_{5\pi X}B^2JJ'\pi'^3 +
    G_{5XX}B^3J'^2\pi'^4 + 16F_{4}B^2JJ'\pi'^3 - 4F_{4X}B^3JJ'\pi'^5, \\
    \end{aligned}
\]
and
\begin{subequations}
\begin{align}
\label{P1}
\mathcal{P}_1 &= \frac{\sqrt{B}}{\sqrt{A}}
\frac{\mbox{d}}{\mbox{d}r}\left[\frac{\sqrt{A}}{\sqrt{B}}
\frac{J^2 \mathcal{H}\left(\mathcal{H} - 2 F_4 B^2 \pi'^4\right)}{2{\cal H} J J' + \Xi \pi'}\right],\\
\mathcal{P}_2 &= \frac{B(A' J - 2 A J')}{A}
\left(2{\cal H} J J' + \Xi \pi'\right)
\; .
\end{align}
\end{subequations}
The expressions for $\mathcal{K}_{12}$ and $\mathcal{K}_{22}$
are given in Appendix C for completeness. According to
eqs.~\eqref{K11} and~\eqref{detK}, both no-ghost
conditions~\eqref{no_ghost} reduce to the following requirement:
\[
\label{no-go}
2{\cal P}_1 - {\cal F} > 0,
\]
where ${\cal F}$ is positive by the stability
conditions in the parity odd sector (see eq.~\eqref{stability_F}).
Importantly, ${\cal P}_1$ in eq.~\eqref{P1} significantly differs from its
analogue in Horndeski theory due to the explicit presence
of the function $F_4$ in the numerator. It is this $F_4$-term that
enables
one to circumvent the no-go theorem.

The general structure of the
no-ghost condition for parity even perturbations is very
similar to the stability condition in non-singular cosmological scenarios
(see Ref.~\cite{Mironov:2018pjk} for discussion).
Indeed, the condition~\eqref{no-go} has the same form as
the central relation in the cosmological
case, with ${\cal P}_1$ being proportional to the derivative
of a certain function $\xi$ ($\dot{\xi}$ in the cosmological case,
see, e.g.,
Ref.~\cite{Mironov:2018oec}). In  complete analogy to the cosmological
setting, the
condition~\eqref{no-go} requires  that ${\cal P}_1$ is  bounded from
below by a positive function:
\[
\label{no-go_alternative}
{\cal P}_1 = \frac{\sqrt{B}}{\sqrt{A}} \xi' > \frac{\mathcal{F}}2,
\]
where
\[
\xi = \frac{\sqrt{A}}{\sqrt{B}}
\frac{J^2 \mathcal{H}\left(\mathcal{H} - 2 F_4 B^2 \pi'^4\right)}{2{\cal H} J J' + \Xi \pi'} \; .
\label{dec10-18-3}
\]
Thus,
$\xi$ must be a monotonously
growing function of radial coordinate and
has to cross zero
somewhere\footnote{A possible loophole is that
  $\mathcal{F}$, and hence $\mathcal{G}$, vanish as
  $r \to -\infty$ and/or $r\to +\infty$, see
  Refs.~\cite{Libanov:2016kfc,Kobayashi:2016xpl,strongcoupling,YOVA}
  for the
  discussion of similar possibility in the
  cosmological context. This is potentially dangerous
  because of possible strong coupling {far}
  away from the wormhole, and
  certainly requires strong deviation from General Relativity there.}.
  In Horndeski theory, the no-go theorem~\cite{Olegi}
is obtained by noting that $\xi$
cannot cross zero in a healthy way (the numerator in
\eqref{dec10-18-3} is manifestly positive for $F_4=0$), which means that the
condition~\eqref{no-go} is inevitably violated at some point and
the solution is plagued by ghosts. The situation in beyond Horndeski
theory
is entirely different:
the extra $F_4$-term in the numerator of $\xi$ makes it possible that
$({\cal H} - 2 F_4 B^2 \pi'^4)$ crosses zero, while ${\cal H}$
remains safely positive, as required by the
stability condition~\eqref{stability_H} in the
parity odd sector. Hence, in beyond Horndeski
theory, the no-ghost condition~\eqref{no-go} may be satisfied
throughout the whole space.

Let us now discuss
radial gradient instabilities. They are absent, provided that
the matrix
$\mathcal{G}_{ij}$ is positive-definite:
\[
\label{no_gradient}
\mathcal{G}_{11}>0,  \qquad \det{\mathcal{G}}>0.
\]
Here
\begin{equation}
\label{G11}
\mathcal{G}_{11}=\frac{4 A B^{2} \left[ \mathcal{G} (\ell+2)(\ell-1)  (2 \mathcal{H} J J'+ \Xi \pi')^2+\ell(\ell+1)(2 J^2 \Gamma \mathcal{H} \Xi \pi'^2  - \mathcal{G} \Xi^2 \pi'^2-4 J^4 \Sigma \mathcal{H}^2/B )\right]}
{ \ell (\ell+1) A^2 \mathcal{H}^2  \left[ 2 J \mathcal{H} \ell (\ell+1)  + \mathcal{P}_2 - 4  J F_4 \ell (\ell+1) B^2 \pi'^4\right]^2},
\end{equation}
\begin{equation}
\label{detG}
\det(\mathcal{G})=\frac{16 A B^3 J'^2 \mathcal{G} (\ell-1) (\ell+2) (2 J^2 \Gamma \mathcal{H} \Xi \pi'^2  - \mathcal{G} \Xi^2 \pi'^2-4 J^4 \Sigma \mathcal{H}^2/B )}
{ \ell (\ell+1) A^2 \mathcal{H}^2  J^2 \pi'^2 \left[2 J \mathcal{H} \ell (\ell+1) +\mathcal{P}_2 -4 J F_4 \ell (\ell+1)  B^2 \pi'^4\right]^2},
\end{equation}
with
\begin{align}
\label{Gamma}
& \Gamma = \Gamma_1 + \frac{A'}{A} \Gamma_2,
\\
& \nonumber\Gamma_{1} = -4XK_{X} + 4\left(G_{4\pi} + 2XG_{4\pi X} + \frac{B\pi'J'}{J}(G_{4X} + 2XG_{4XX}) - \frac{BJ'}{J}\pi'(G_{5\pi} + XG_{5\pi X})\right)
\\& - \frac{16J'}{J}B\pi'X(2F_{4} + XF_{4X}), \nonumber\\
&\Gamma_{2} = 2B\pi'\left(G_{4X} - B\pi'^2G_{4XX} - G_{5\pi} - XG_{5\pi X} - \frac{BJ'\pi'}{2J}(3G_{5X} + 2XG_{5XX})\right)
\nonumber\\& - 8B\pi'X(2F_{4} + XF_{4X}),
\nonumber
\end{align}
and
\begin{align}
& \Sigma = XF_{X} + 2F_{XX}X^2 - B\pi'\left(\frac{4J'}{J } + \frac{ A'}{A} \right)X(K_{X} + XK_{XX}) - 2XK_{\pi} - 4K_{\pi X}X^2 + 2X^2K_{\pi X} +
\nonumber\\&
 2\left(\frac{1 - BJ'^2}{J^2} - \frac{BJ'}{J} \frac{A'}{A}\right)X(G_{4X} + 2XG_{4XX}) -
      \frac{4BJ'}{J}\left( \frac{J'}{J } + \frac{A'}{A}\right)X^2(3G_{4XX} + 2XG_{4XXX}) +
      \nonumber\\&
       2B\pi'\left(\frac{4J'}{J } + \frac{ A'}{A} \right)X( \frac{3}{2} G_{4\pi X} + XG_{4 \pi XX}) - \frac{B\pi'(1 - 3BJ'^2)}{J^2} \frac{A'}{A} X(G_{5X} + XG_{5XX}) +
       \nonumber\\&
      \frac{2B^2J'^2\pi'}{J^2} \frac{A'}{A}\left(2X^2G_{5XX} + X^3G_{5XXX}\right) - 2\left(\frac{1 - BJ'^2}{J^2 } - \frac{BJ'}{J} \frac{A'}{A} \right)X(G_{5\pi} + 2XG_{5\pi X})
       \nonumber\\&
        + 2\frac{BJ'}{J}\left( \frac{J'}{J } + \frac{A'}{A}\right)X^2(3G_{5\pi X} + 2XG_{5\pi XX}) + \frac{2}{J^2}X^2G_{5\pi X}
      \nonumber\\&
      + \frac{B^3J'(A'J + AJ')}{AJ^2}\pi'^4(12F_{4} - 9F_{4X}B\pi'^2 + F_{4XX}B^2\pi'^4).
\label{KSI}
\end{align}
The expressions for $\mathcal{G}_{12}$ and $\mathcal{G}_{22}$ are
given in  Appendix C. Both conditions~\eqref{no_gradient}
are satisfied provided that
\[
\label{P3}
2 J^2 \Gamma \mathcal{H} \Xi \pi'^2  - \mathcal{G} \Xi^2 \pi'^2-4 J^4 \Sigma \mathcal{H}^2/B  > 0.
\]
The sound speeds squared of the even-parity modes along the radial
direction are equal
to the eigenvalues of the matrix $(AB)^{-1}(\mathcal{K})^{-1}\mathcal{G}$:
\begin{equation}
\label{speed_even}
c_{s1}^2 = \frac{\mathcal{G}}{\mathcal{F}},
\qquad
c_{s2}^2 = \frac{(2 J^2 \Gamma \mathcal{H} \Xi \pi'^2  - \mathcal{G} \Xi^2 \pi'^2-4 J^4 \Sigma \mathcal{H}^2/B )}
{\left(2{\cal H} J J' + \Xi \pi'\right)^2 (2{\cal P}_1-{\cal F})} \; .
\end{equation}
Note that $c_{s1}^2$ coincides with the radial
speed squared of the odd parity mode $c_r^2$ in
eq.~\eqref{speed_odd}, which enables one to interpret
it as the radial propagation speed of two tensor degrees of freedom.

Thus, the parity even modes have neither ghosts nor
radial gradient instabilities provided that the conditions
\eqref{no-go},~\eqref{P3} are satisfied. These modes are
subluminal along the radial direction
when $c_{s1}, c_{s2} \le 1$. We will see in the next
section that all these constraints can indeed be satisfied.

As we pointed out in section~\ref{sec:intro}, our stability analysis
(like the ones in Refs.~\cite{Trincherini,Kobayashi:even})
is incomplete, as we do not study angular gradient instabilities
{as well as ``slow''
tachyonic}
instabilities in the parity even sector.
In other words, the
stability conditions associated with matrices $\mathcal{M}_{ij}$
and $\mathcal{Q}_{ij}$ in the action~\eqref{action_even}
are yet to be addressed.
This
issue,  technically quite challenging, is left for the future.


\section{Wormhole beyond Horndeski: an example}
\label{sec:wormhole}
Let us give a specific example of beyond
Horndeski theory admitting a wormhole solution which is free of
 {all  ghost and gradient instabilities in parity odd sector} and
 free of ghosts and radial gradient instabilities in parity even sector.
Similarly to the
cosmological setting, we adopt a ``reconstruction'' approach.
Namely, we arbitrarily choose
the background metric of wormhole form,
and cook up the Lagrangian functions in such a way that the
equations for background and
stability conditions~\eqref{stability_F}-\eqref{stability_H},
~\eqref{no-go} and~\eqref{P3} are satisfied. At the same time
we ensure that the sound speeds in~\eqref{speed_odd}
and~\eqref{speed_even} are at most luminal. The whole procedure
is very similar to that carried out for the bouncing solution in
Ref.~\cite{Mironov:2018oec}, modulo the form of asymptotics.
Here we require only that the space-time is asymptotically flat
 and do not impose any restrictions on the asymptotic
 behavior of the Lagrangian.

 We begin with choosing the
 same
 specific form of the metric
functions in~\eqref{backgr_metric} {as in Refs.~\cite{Ellis,Bronnikov}}:
\[
\label{ABJ}
A = B = 1, \qquad J = \sqrt{r^2 + \tau^2},
\]
where the parameter $\tau \equiv R_{min}$
regulates the size of the wormhole
throat at $r=0$. Equation~\eqref{ABJ} immediately implies that
gravity is modified, as compared to General Relativity, even
far away from the wormhole: the wormhole mass vanishes, and yet
the metric is not exactly flat at large $|r|$.

The scalar field, supporting the wormhole, is also
static and spherically symmetric:
\[
\label{galileon}
\pi (r) = c_1 \cdot \mbox{arcsinh}\left(\frac{r}{r_0}\right), \qquad
X = -\frac{c_1^2}{2\:r_0^2\left(1+\frac{r^2}{r_0^2}\right)}.
\]
For monotonously growing scalar field background, the solution
$\pi(r)$ can be always transformed into the form~\eqref{galileon}
by field redefinition.
In what follows we choose
\[
c_1 = r_0 = \tau,
\]
although this choice is not obligatory.

To find the Lagrangian,  we choose the
following Ansatz:
\begin{subequations}
\label{ansatz}
\begin{align}
\label{F}
& F(\pi, X) = f_0(\pi) + f_1(\pi)\cdot X + f_2(\pi)\cdot X^2, \\
\label{G4}
& G_4(\pi, X) = \frac12 + g_{40}(\pi) + g_{41}(\pi) \cdot X,\\
\label{F4}
& F_4(\pi, X) = f_{40}(\pi) + f_{41}(\pi) \cdot X,
\end{align}
\end{subequations}
while $K(\pi, X) = 0$, $G_5(\pi, X) = 0$ and $F_5(\pi, X) = 0$
in full analogy with the bouncing setup of
Ref.~\cite{Mironov:2018oec}.

Our main requirement is that {there are no ghost and gradient
instabilities of the types considered in this paper in
both parity even and parity odd perturbation sectors about the wormhole
set-up~\eqref{ABJ}.
To this end,} we make use of
constraints~\eqref{stability_G}--\eqref{stability_H},
~\eqref{sublu-odd-1}--\eqref{sublu-odd-2},
~\eqref{no_ghost},
\eqref{no_gradient} and also $c_{s2}\le 1$, where $c_{s2}^2$ is given by
\eqref{speed_even}.
Let us recall
the no-ghost condition~\eqref{no_ghost} for the parity even
modes:
\[
\label{D:no_go}
\frac{\sqrt{B}}{\sqrt{A}}\cdot \frac{\mbox{d}}{\mbox{d}r}\left[ \frac{\sqrt{A}}{\sqrt{B}}
\frac{J^2 \mathcal{H}\left(\mathcal{H} - 2 F_4 B^2 \pi'^4\right)}{2{\cal H} J J' + \Xi \pi'} \right] > \frac{{\cal F}}{2},
\]
where we use the definition~\eqref{P1} of
${\cal P}_1$. Making use of~\eqref{ansatz}, we express ${\cal H}$,
${\cal F}$, ${\cal G}$, $\Xi$, $\Sigma$ and $\Gamma$
in terms of $g_{4i}$, $f_{4i}$ and $f_{j}$:
\begin{subequations}
\begin{align}
\label{D:H}
& {\cal H} = {\cal G} = 1 + 2 g_{40}(\pi)+g_{41}(\pi)\cdot \pi'^2 + 
2 f_{40}(\pi)\cdot \pi'^4-f_{41}(\pi)\cdot \pi'^6,\\
\label{D:F}
& {\cal F} = 1+2 g_{40}(\pi)-g_{41}(\pi)\cdot\pi'^2, \\
\label{D:Xi}
& \Xi = 2\: (r^2+\tau^2)\cdot g_{40}'(\pi)+\pi' \cdot[ 4 r g_{41}(\pi)
+ 4 r \pi'^2\cdot(4 f_{40}(\pi)- 3 f_{41}(\pi)\cdot \pi'^2) 
\\\nonumber
&- 3 \pi' (r^2+\tau^2)\cdot g_{41}'(\pi)],\\
\label{D:Sigma}
& \Sigma = -\frac{\pi'^2}{2(r^2 + \tau^2)^2} 
[(r^2+\tau^2)^2 \cdot f_1(\pi) + 2 \tau^2 \cdot g_{41}(\pi)
+ 12 \pi' r (r^2+\tau^2)\cdot g_{41}'(\pi)\\\nonumber
& - 3 \pi'^2 ( (r^2+\tau^2)^2 \cdot f_2(\pi)
+ 8 r^2 \cdot  f_{40}(\pi)- 10 r^2  f_{41}(\pi) \cdot \pi'^2)],\\
\label{D:Gamma}
& \Gamma = \frac{2 \pi'}{r^2 + \tau^2}
[2 r \cdot g_{41}(\pi) - 3 \pi' (r^2+\tau^2) \cdot g_{41}'(\pi) + 
8 r\pi'^2 \cdot f_{40}(\pi)  - 6 r \pi'^4 \cdot f_{41}(\pi) ] \\\nonumber
&+4 g_{40}'(\pi).
\end{align}
\end{subequations}
Here and in what follows we keep 
$\pi$ in the argument of functions to make 
the expressions easy to read. However, we have 
chosen the coordinate dependence of $\pi$ in eq.~\eqref{galileon}, 
so, in fact we work with functions of $r$.
Note that ${\cal H} = {\cal G}$ due to our choice
$G_5(\pi,X) = 0$
(see eqs.~\eqref{cal_h} and~\eqref{cal_g}).
To avoid superluminal propagation,
we set for simplicity
\[
\label{D:HF_choice}
{\cal H} = {\cal G} ={\cal F} = 1,
\]
which immediately gives $c_r^2 = c_{\theta}^2 = c_{s1}^2 = 1$.
The choice in eq.~\eqref{D:HF_choice} together with
relations~\eqref{D:H} and~\eqref{D:F} enables one to express
the functions
$g_{41}(\pi)$ and $f_{40}(\pi)$ through $g_{40}$ anf $f_{41}$:
\begin{subequations}
\begin{align}
\label{D:g41}
& g_{41}(\pi) = 2 g_{40}(\pi)\cdot \mbox{cosh}^2\left(\frac{\pi}{\tau}\right),\\
\label{D:f40}
& f_{40}(\pi) = \frac12 \; f_{41}(\pi) \cdot \mbox{sech}^2
\left(\frac{\pi}{\tau}\right) - 2 g_{40}(\pi) \cdot
\mbox{cosh}^4 \left(\frac{\pi}{\tau}\right) .
\end{align}
\end{subequations}
As we discussed in section~\ref{quad_action_even}, in order to circumvent the
no-go theorem one has to make sure that
$\left(\mathcal{H} - 2 F_4 B^2 \pi'^4\right)$ in
eq.\eqref{D:no_go} crosses zero at some point. Since we have
fixed ${\cal H}$ in eq.\eqref{D:HF_choice}, let us choose
$F_4(\pi,X)$ as follows:
\[
\label{D:F4_choice}
F_4(\pi,X) \equiv f_{40}(\pi) + f_{41}(\pi)\cdot X =
w\cdot \mbox{sech}\left(\frac{r}{\tau}+u\right),
\]
where $w$ and $u$ are parameters and $\tau$ still defines the
size of the wormhole throat. Again
$f_{40}$ and $f_{41}$ are in fact
functions of $r$, see eq.~\eqref{galileon}. 
For sufficiently large $w$, the above
choice  leads to the
required zero-crossing of the numerator in eq.~\eqref{D:no_go}.
In our numerical examples below we set
\[
w = 1, \qquad u = \frac{1}{10}, \qquad \tau = 10.
\]
Note that for $u\neq 0$, our set-up is not invariant under
reflection $r \to -r$. We introduced the parameter $u$
into our Ansatz to emphasize that there is no fine-tuning
(while with $u=0$ one has $F_4' = 0$ at the wormhole throat $r=0$).

Making use of eqs.~\eqref{D:g41} and~\eqref{D:f40}, we obtain both
$g_{40}(\pi)$ and $g_{41}(\pi)$ explicitly from eq.~\eqref{D:F4_choice}:
\[
\label{D:g40}
g_{40}(\pi)= \frac12\; g_{41}(\pi) \cdot  
\mbox{sech}^2\left(\frac{\pi}{\tau}\right) =
-\frac{w}2 \cdot \mbox{sech}\left[u+\mbox{sinh}\left(\frac{\pi}{\tau}\right)\right]
\cdot \mbox{sech}^4\left(\frac{\pi}{\tau}\right).
\]
Hence, we have completely defined $G_4(\pi,X)$.

Now we turn to the denominator of the no-ghost
condition~\eqref{D:no_go}, where $\Xi$ involves the yet undetermined
function $f_{41}(\pi)$, see eq.~\eqref{D:Xi}.
We choose the function $f_{41}(\pi)$ is such a way that the
denominator of the inequality~\eqref{D:no_go} behaves as follows
(except for the vicinity of $r=0$):
\[
\label{D:xi_denominator}
(2 {\cal H} J J' + \Xi \pi') \approx  {\cal H} J J' \; ,
\]
which is sufficient for satisfying the no-ghost
condition~\eqref{D:no_go} everywhere. The approximate
equality~\eqref{D:xi_denominator}
holds for the function $f_{41}(\pi)$ given by
\[
\begin{aligned}
\label{D:f41}
&f_{41}(\pi)= \frac14 \cosh^2\left(\frac{\pi}{\tau}\right) 
\cdot \left(
\cosh^4\left(\frac{\pi}{\tau}\right) + 10 w \cdot 
\mbox{sech}\left[ u + \sinh\left(\frac{\pi}{\tau}\right)\right]\right.\\
&\left. - 2 w \cdot \cosh\left(\frac{\pi}{\tau}\right)
\cdot \coth\left(\frac{\pi}{\tau}\right)
\cdot \mbox{sech}\left[ u + \sinh\left(\frac{\pi}{\tau}\right)\right]
\cdot \mbox{tanh}\left[\sinh \left(\frac{\pi}{\tau}\right)\right]\right).
\end{aligned}
\]
Note that $f_{41}(\pi)$ is non-singular at $r=0$, where $\pi=0$.
Thus, we have chosen $g_{40}(\pi)$, $g_{41}(\pi)$,
$f_{40}(\pi)$ and $f_{41}(\pi)$ in such a way
that the constraints~\eqref{stability_F}--\eqref{stability_H} and~\eqref{no_ghost} are satisfied.

The functions yet undefined are $f_{0}(\pi)$, $f_{1}(\pi)$ and
$f_{2}(\pi)$. We find $f_{0}(\pi)$ in terms
of $f_{1}(\pi)$ and $f_{2}(\pi)$ by solving the background equation
${\cal E}_A = 0$:
\[
\begin{aligned}
\label{D:f0}
&f_0(\pi)= \frac14 \cdot \mbox{sech}^2\left(\frac{\pi}{\tau}\right)
\left[2 f_1(\pi) - f_2(\pi) \cdot 
\mbox{sech}^2\left(\frac{\pi}{\tau}\right) \right. \\
& \left.
+ \frac{1}{\tau^2}\cdot \mbox{sech}^2\left(\frac{\pi}{\tau}\right)
\left(4 + 8 w\cdot \mbox{sech}\left(\frac{\pi}{\tau}\right) \cdot
\mbox{sech}\left[ u + \sinh\left(\frac{\pi}{\tau}\right)\right]\cdot
\left[- 5\cdot \mbox{sech}\left(\frac{\pi}{\tau}\right) \right.\right.\right.\\
&\left.\left.\left.
+ 6 \cdot \mbox{sech}^3\left(\frac{\pi}{\tau}\right)
- \mbox{tanh}\left(\frac{\pi}{\tau}\right) \cdot 
\mbox{tanh}\left[ u + \sinh\left(\frac{\pi}{\tau}\right)\right]\right]\right)\right].
\end{aligned}
\]
In the same manner, $f_{1}(\pi)$ is found from the background equation
${\cal E}_B = 0$ with $f_{0}(\pi)$ substituted from eq.~\eqref{D:f0}. 
Thus, the solution for $f_{1}(\pi)$ involves only $f_{2}(\pi)$ and reads:
\[
\begin{aligned}
\label{D:f1}
&f_1(\pi)= \frac{1}{4\: \tau^2}\;\mbox{sech}^2\left(\frac{\pi}{\tau}\right) \left[4 \tau^2 \cdot f_2(\pi) - 7 - \cosh\left(\frac{2 \pi}{\tau}\right)
- 4 w \cdot \mbox{sech}\left(\frac{\pi}{\tau}\right) \cdot
\mbox{sech}\left[ u + \sinh\left(\frac{\pi}{\tau}\right)\right]
\right. \\
&\left.
\times 
\mbox{tanh}\left(\frac{\pi}{\tau}\right) \cdot
\left( 2\: \mbox{sech}\left(\sinh\left(\frac{\pi}{\tau}\right)\right) \cdot
\mbox{sech}\left[ u + \sinh\left(\frac{\pi}{\tau}\right)\right] \cdot 
\sinh(u) - 9 \: \mbox{sech}\left(\frac{\pi}{\tau}\right) \cdot 
 \mbox{tanh}\left(\frac{\pi}{\tau}\right)
\right.\right.\\
&\left.\left.
  + \mbox{tanh}\left[\sinh\left(\frac{\pi}{\tau}\right)\right]\right)\right].
\end{aligned}
\]
Then the background equation ${\cal E}_J = 0$ is satisfied
due to our specific choice of metric functions in eq.~\eqref{ABJ}
and ${\cal E}_\pi = 0$ is valid automatically.

Now, we still have to ensure that there are no radial
gradient instabilities,
i.e. inequality~\eqref{no_gradient} holds. An additional constraint
is imposed by the requirement that $c_{s2}^2 \leq 1$ in
eq.~\eqref{speed_even}.  These requirements are satisfied by
a judicial choice of
the remaining function $f_{2}(\pi)$
entering $\Sigma$ (see eq.~\eqref{D:Sigma}). Indeed, the
conditions~\eqref{no_gradient} and \eqref{speed_even}
boil down to
\[
\label{D:constraint_f2}
0 < 2 J^2 \Gamma \mathcal{H} \Xi \pi'^2  - \mathcal{G} \Xi^2 \pi'^2-4 J^4 \Sigma \mathcal{H}^2/B \leq (2 {\cal H} J J' + \Xi \pi' )^2 (2 {\cal P}_1 -{\cal F}).
\]
We have already ensured that the right hand side here is
positive, so the latter inequalities can be satisfied by
choosing $\Sigma$, and hence $f_2$ an an appropriate way.
As an example, we choose $\Sigma(r)$ in such a way that
\[
\label{D:for_sigma}
(2 J^2 \Gamma \mathcal{H} \Xi \pi'^2  - \mathcal{G} \Xi^2 \pi'^2-4 J^4 \Sigma \mathcal{H}^2/B)\cdot J^{-2} = \frac{2 r^2- 3\: r \tau \cdot \tanh(r)+
3\tau^2}{2(r^2+\tau^2)}.
\]
One can check that the inequality~\eqref{D:constraint_f2} is indeed
satisfied by this choice. The corresponding function $f_2$ is given in Appendix D.

\begin{figure}[H]\begin{center}\hspace{-1cm}
{\includegraphics[width=0.5\textwidth]{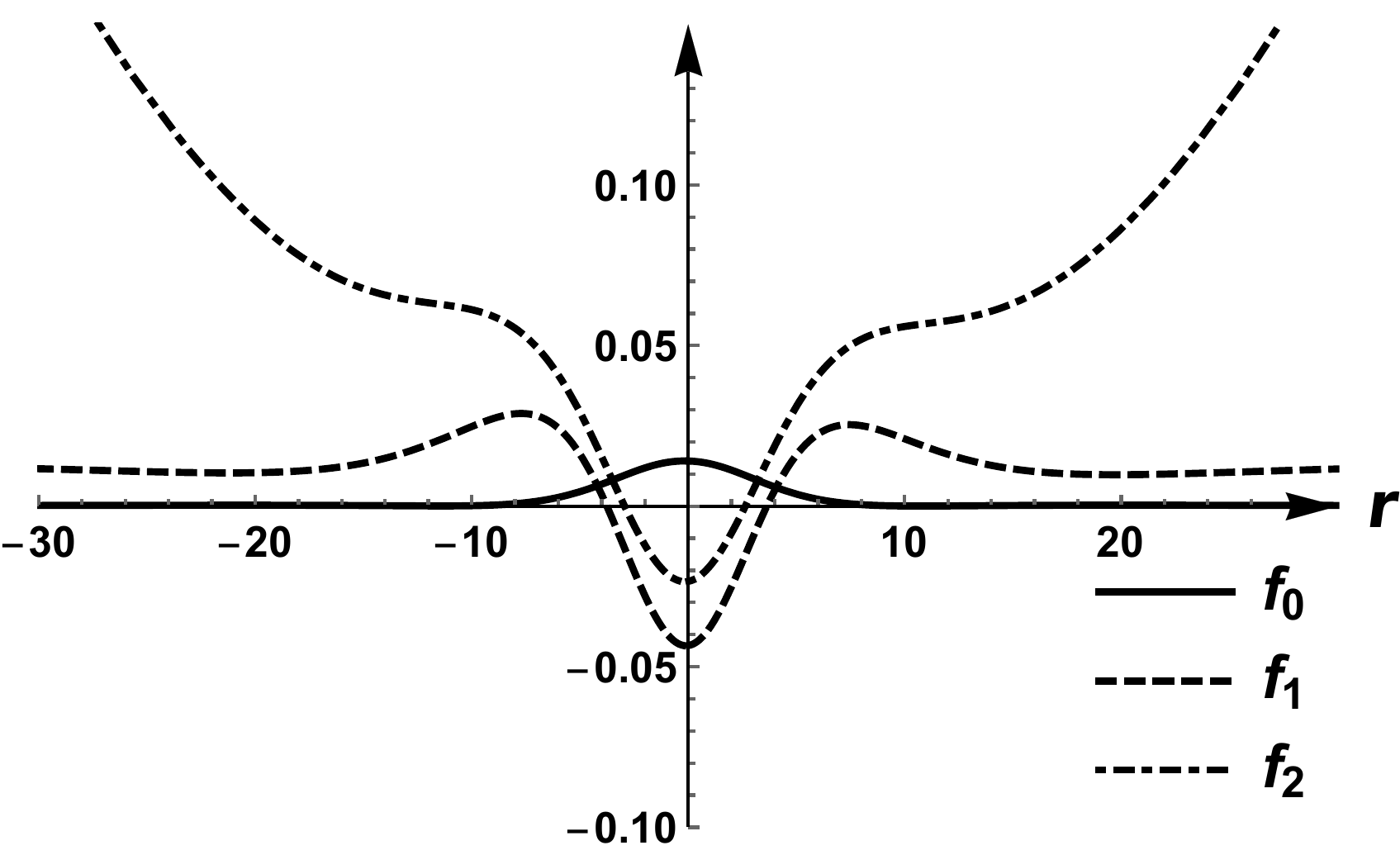}}\hspace{2.8cm}\hspace{-3cm}
{\includegraphics[width=0.5\textwidth] {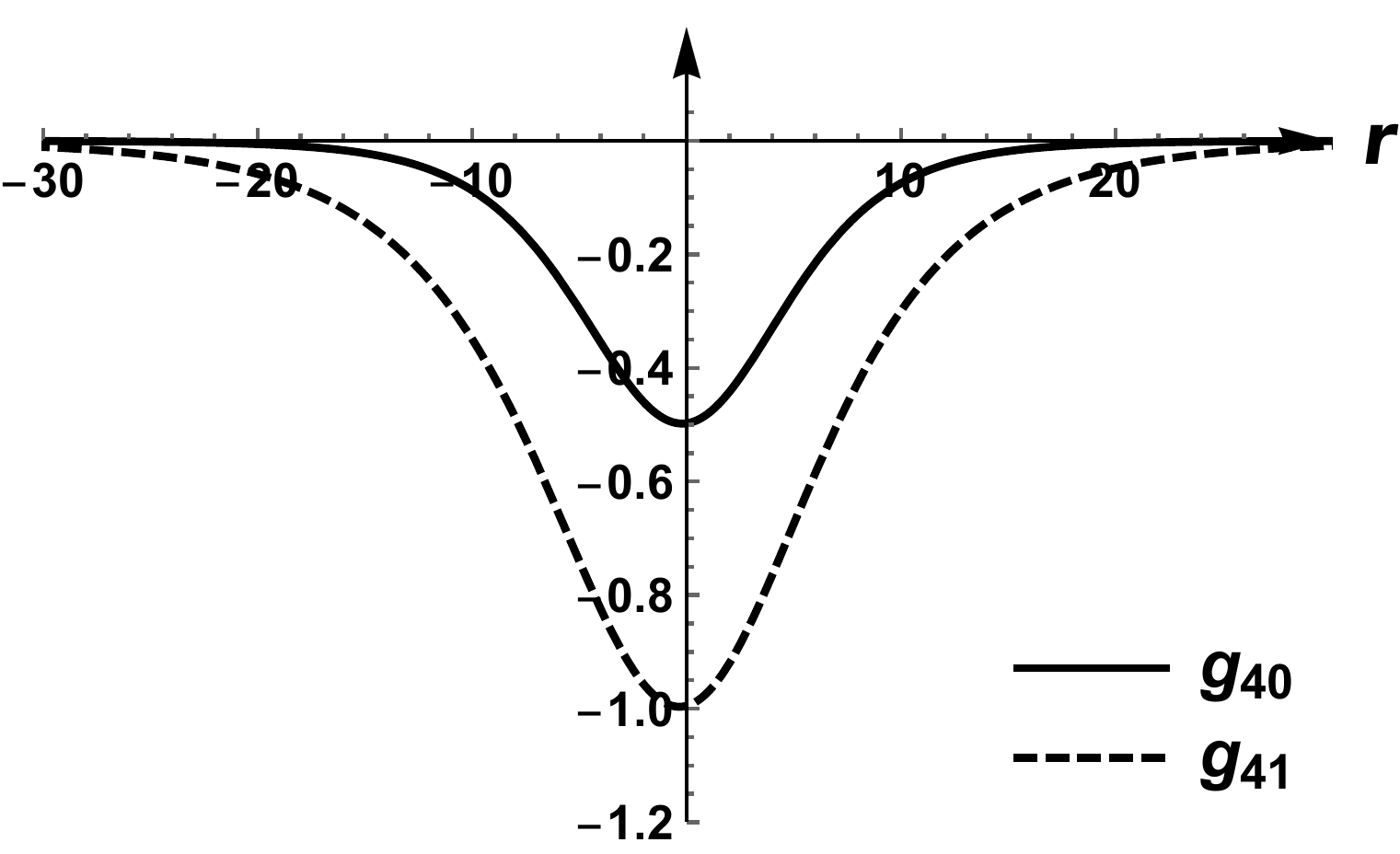}}

\vspace{-0.2cm}
{\includegraphics[width=0.5\textwidth]
{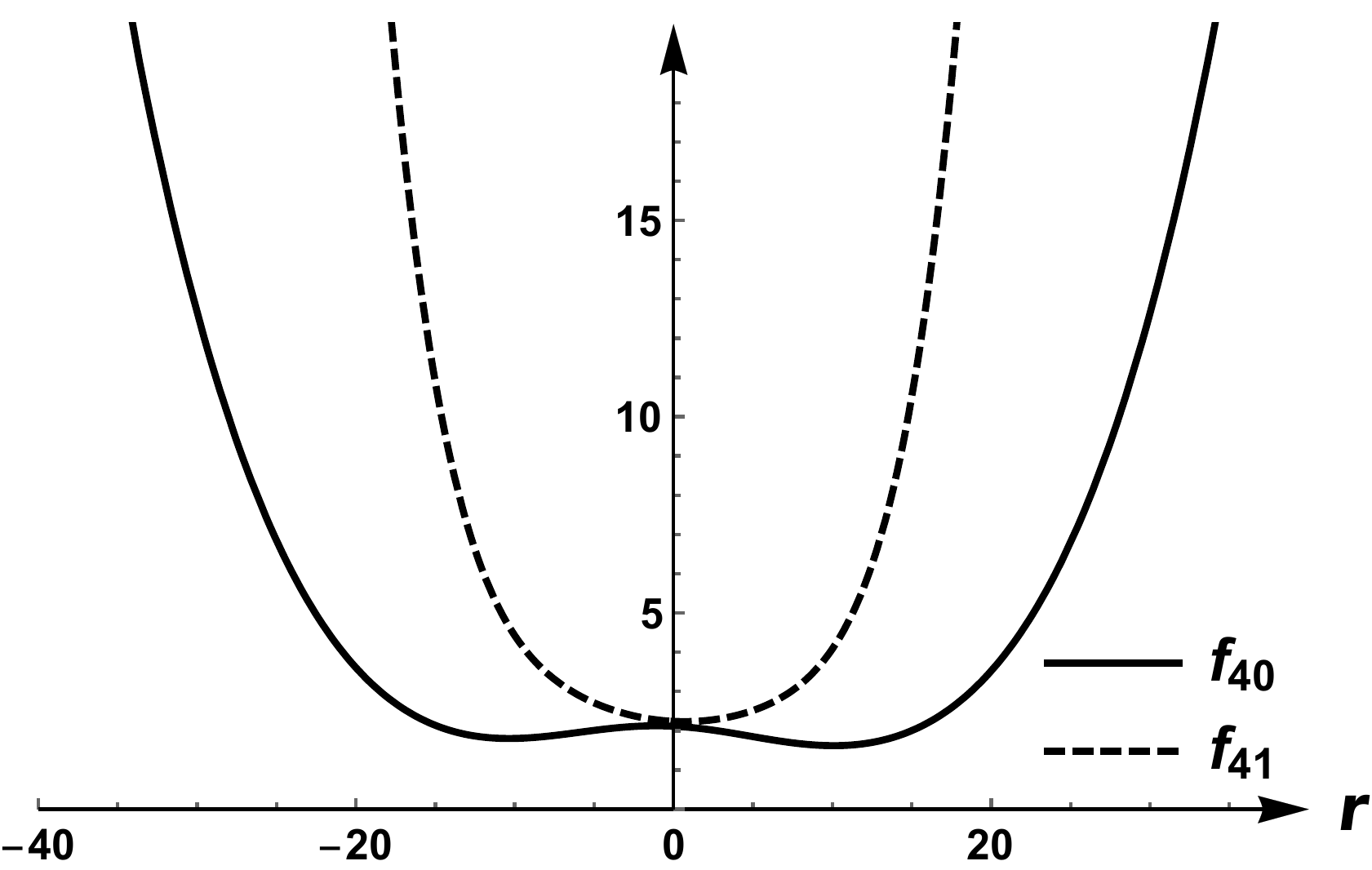}}\hspace{1cm}
\caption{\footnotesize{The Lagrangian functions $f_0(r)$, $f_1(r)$, $f_2(r)$, $g_{40}(r)$, $g_{41}(r)$,
    $f_{40}(r)$ and $f_{41}(r)$, with the following choice of the parameters: $u=1/10$, $w=1$, $r_0=10$, $c_1=10$, and $\tau = 10$. This choice
    guarantees that the size
    of the wormhole throat safely exceeds the Planck length.}} \label{f-g}
\end{center}\end{figure}

The reconstructed functions entering
eqs.~\eqref{F}--\eqref{F4} are shown in Fig.\ref{f-g}
They
have the following asymptotic
behaviour as $r \rightarrow \pm \infty$:
\[
\label{function_asymp}
f_{0}\propto \frac{1}{r^2},\quad f_{1} \propto \mbox{const}, \quad
f_{2}\propto r^2, \quad
g_{40} = g_{41} \propto e^{-r}, \quad
f_{40}  \propto r^4,\quad f_{41}  \propto r^6 \; ,
\]
which results in the following asymptotics of $F(\pi,X)$, $G_4(\pi,X)$ and
$F_4(\pi,X)$ far away from the throat, where $|\pi(r)| = c_1 \log|r|$:
\[
\label{asymptotic_Lagrangian}
\begin{aligned}
&F(\pi, X) = q_1\cdot\mbox{sinh}^{-2}(\pi) + q_2\cdot X + q_3\cdot \mbox{sinh}^2(\pi)\cdot X^2, \\
&G_4(\pi, X) = \frac12 , \\
&F_4(\pi, X) = q_4\cdot \mbox{sinh}^4(\pi) + q_5\cdot \mbox{sinh}^6(\pi) \cdot X,
\end{aligned}
\]
where $q_i$, $i=1,\dots,5$ are positive coefficients,
whose values are not important for us.
Even though the asymptotic value of $G_4$ is that of
General Relativity, {weak gravity regime is grossly different from
General Relativity even asymptotically.}
Indeed,
non-vanishing $F_4(\pi, X)$ gives
a non-trivial contribution to the sound speeds of tensor-like modes in eqs.~\eqref{speed_odd}.

\begin{figure}[H]\begin{center}
{\includegraphics[width=0.7\textwidth]
{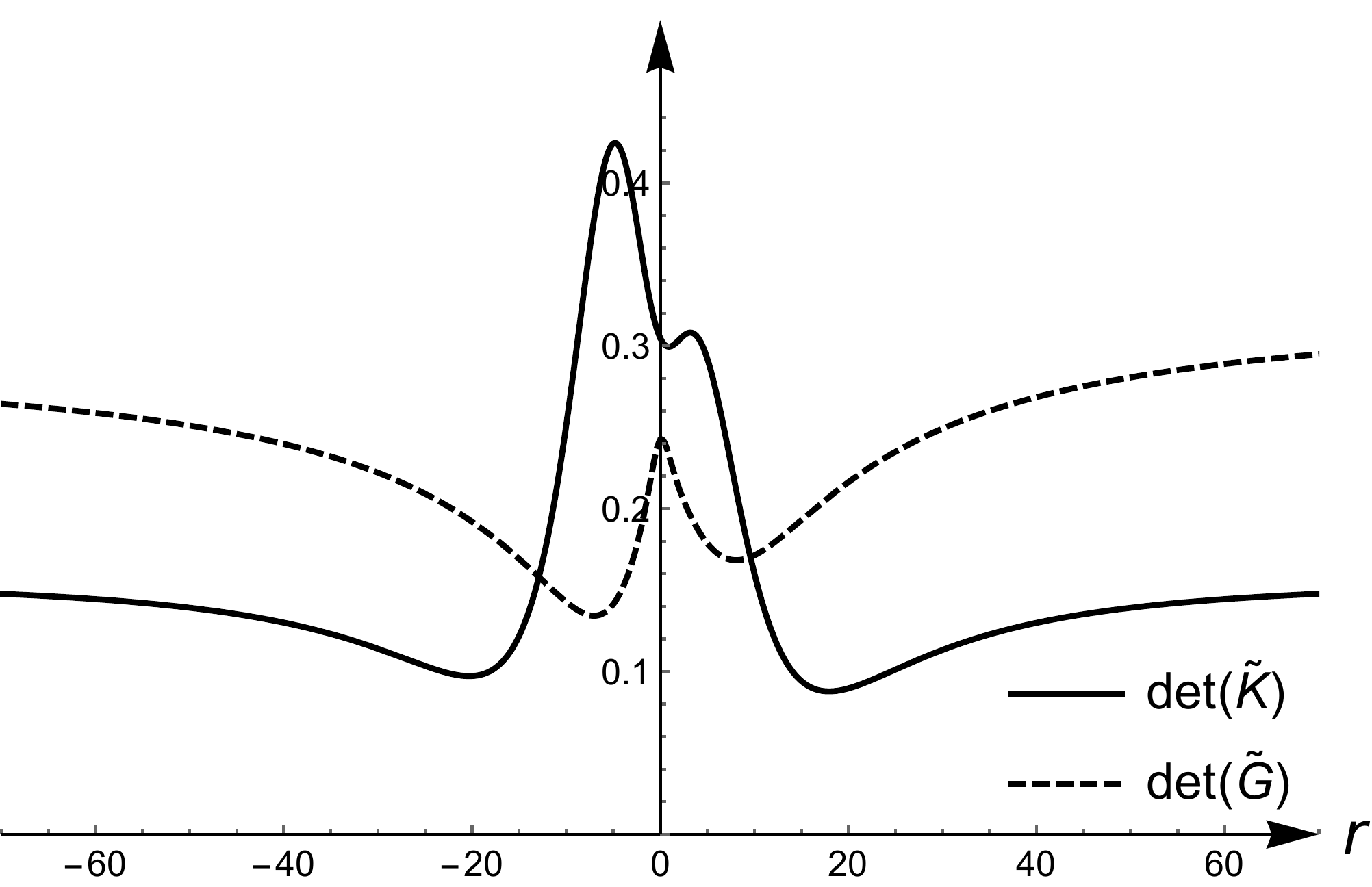}}\vspace{1.5cm}
{\includegraphics[width=0.7\textwidth]
{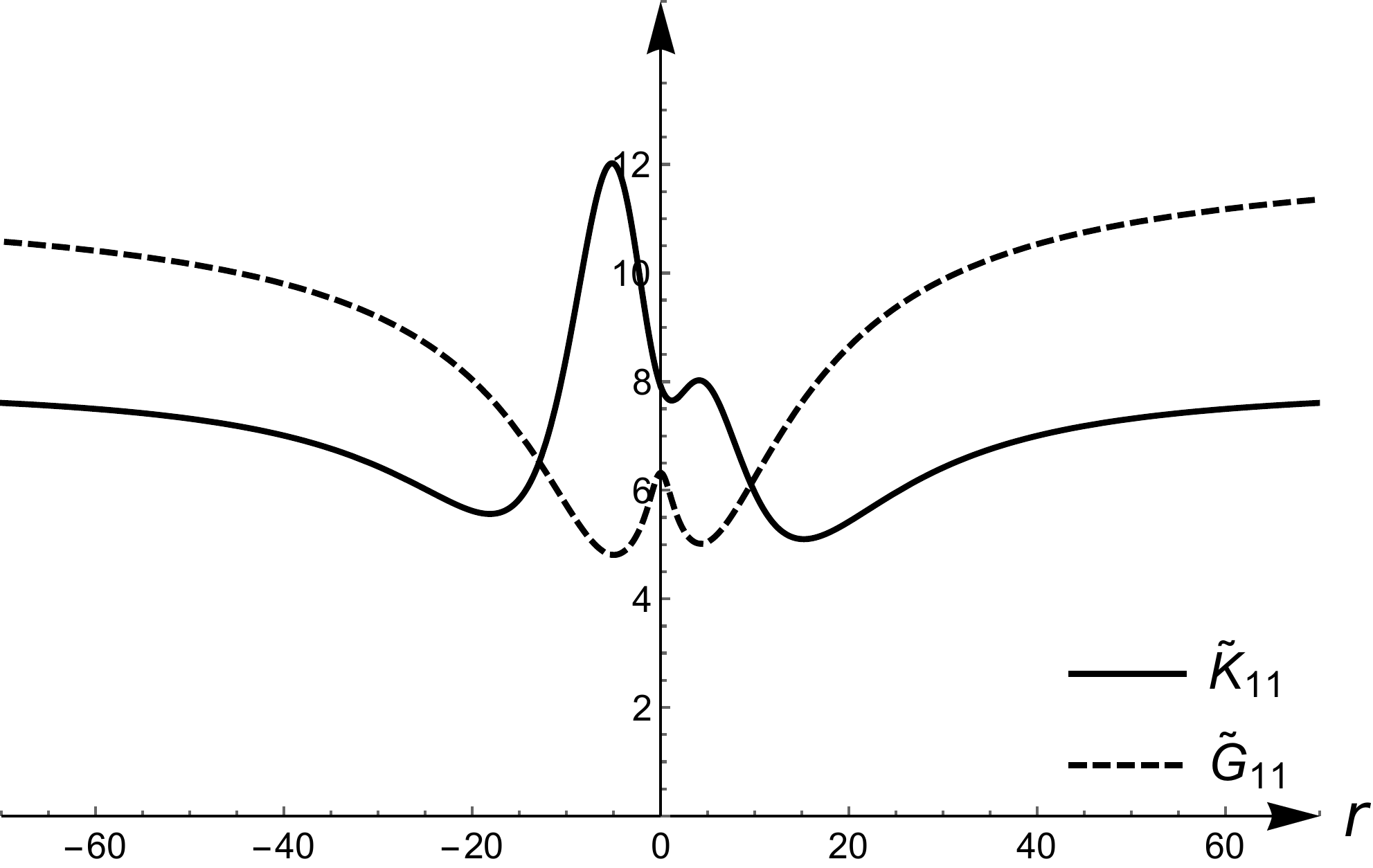}}
\caption{\footnotesize{Functions $\tilde{\cal K}_{11}$, 
$\det(\tilde{\cal K})$, $\tilde{\cal G}_{11}$ and 
$\det(\tilde{\cal G})$, which govern the stability of the 
parity even sector (we choose $\ell = 10$ here for definiteness). 
}}
\label{KG}
\end{center}\end{figure}

Let us demonstrate explicitly that our solution satisfies
our subset of stability conditions.
We have arranged the Lagrangian functions so
that ${\cal H} = {\cal G} = {\cal F} = 1$, which automatically
ensures that the parity odd
modes {obey}~\eqref{stability_G}-\eqref{stability_H} and \eqref{sublu-odd-1},
\eqref{sublu-odd-2}.
As for the parity
even modes, it follows from eqs.~\eqref{K11},~\eqref{detK},~\eqref{G11}
and~\eqref{detG} that {the absence of ghosts and gradient
instabilities in radial direction}
amounts to satisfying the following inequalities:
\begin{gather}
\label{asympK}
 \tilde{\cal K}_{11} = \frac{\left( 2{\cal H} J J' +
\Xi \pi' \right)^2 \left[ \ell (\ell+1){\cal P}_1-{\cal F}\right]}
{J^2} > 0, \\
 \det(\tilde{\cal K}) =  \frac{\left( 2{\cal H} J J' +
\Xi \pi' \right)^2 \left[ 2{\cal P}_1-{\cal F}\right]}
{\pi'^2 J^4} > 0,\\
 \tilde{\cal G}_{11} = \frac{\left[ \mathcal{G} (\ell+2)(\ell-1)  (2 \mathcal{H} J J'+ \Xi \pi')^2+\ell(\ell+1)(2 J^2 \Gamma \mathcal{H} \Xi \pi'^2  - \mathcal{G} \Xi^2 \pi'^2-4 J^4 \Sigma \mathcal{H}^2/B )\right]}
{ J^2} > 0, \\
\label{asympG}
 \det(\tilde{\cal G}) = \frac{(2 J^2 \Gamma \mathcal{H} \Xi \pi'^2  - \mathcal{G} \Xi^2 \pi'^2-4 J^4 \Sigma \mathcal{H}^2/B )}
{ \pi'^2 J^4} > 0.
\end{gather}
Here we simplified expressions for ${\cal K}_{11}$, etc., and
introduced the factor $J^{-2}$ to match the asymptotics:
$\tilde{\cal K}_{11} = {\cal K}_{11}$ as $r \rightarrow \pm \infty$, etc.
The functions~\eqref{asympK}--\eqref{asympG} are shown in Fig.\ref{KG}.

\begin{figure}[H]\begin{center}\hspace{-1cm}
{\includegraphics[width=0.7\textwidth]
{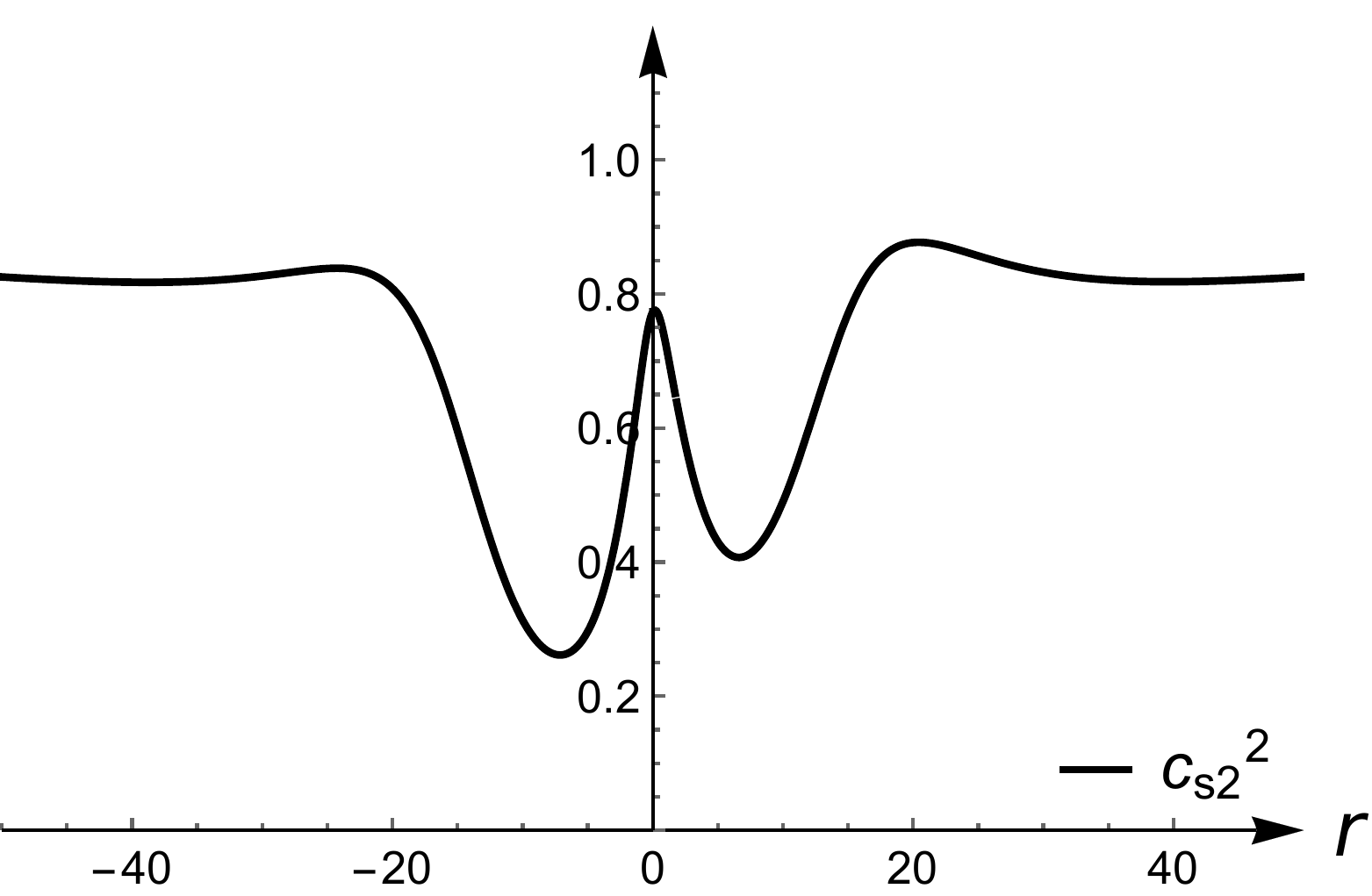}}\hspace{-1cm}
\caption{\footnotesize{Sound speed squared $c_{s2}^2$ in the parity even sector. The choice
of parameters is the same as in Fig.\ref{f-g}:
$u=1/10$, $w=1$, $r_0=10$, $c_1=10$, and $\tau = 10$.}}
\label{cs}
\end{center}\end{figure}

The corresponding sound speed squared $c_{s2}^2$ for parity
even modes~\eqref{speed_even} is shown
in Fig.\ref{cs} (recall that another sound speed $c_{s1}^2$
in parity even sector is
equal to 1).

Thus, parity odd perturbations around the constructed wormhole solution
are {healthy against ghosts and gradient instabilities},
while parity even sector is free of ghosts
and radial gradient instabilities.
As we already mentioned in section~\ref{sec:intro},
 our solution is not completely satisfactory
due to vanishing mass of the wormhole (see eq.~\eqref{ABJ})
and non-trivial asymptotics,{with gravity deviating strongly from
General Relativity as $|r| \to \infty$
even in the weak field limit.
Nevertheless, our findings 
suggest that beyond Horndeski theories may admit stable, and possibly
phenomenologically viable wormholes.}

{
\section{Conclusions}
\label{sec:conclusion}

To summarize,
generalizations of Horndeski theories known as ``beyond Horndeski''
are interesting candidates for  theories admitting stable Lorentzian
wormholes. Indeed, this work, together with
the independent analysis of Ref.~\cite{Trincherini} and 
our earlier
study~\cite{Mironov:2018pjk}
has shown that beyond Horndeski theories circumvent the no-go theorem
forbidding the existence of stable, static, spherically symmetric wormholes
in Horndeski theories, the latter being a fairly general class of
scalar-tensor gravities. Moreover, we have seen that
ghost and gradient instabilities
may be absent both in the parity odd sector of perturbations and
among parity even perturbations travelling along radial direction. 
A particularly new point of this paper is the
construction of an explicit example of such a wormhole. It remains to be
seen whether one can construct completely stable wormholes in
beyond Horndeski theories, and if so, whether 
the wormholes of this sort are viable in the sense that
physics 
in the large distance asymptotics is consistent with
experimental and observational tests.}

 \section*{Acknowledgements}{}
 The authors are grateful to Evgeny Babichev and
 Sergei Sibiryakov
 for helpful discussions
 and Enrico Trincherini for
 useful correspondence. {We are indebted to the anonymous referees
 for instructive criticism.}
The work of S.M. and V.V. was supported in part by the RFBR grant
18-32-00812 and the Foundation for the Advancement of
Theoretical Physics and Mathematics “BASIS” grant.

\section*{Appendix A}{}
In this Appendix we collect the Einstein and Galileon equations
of motion derived for beyond Horndeski theory~\eqref{lagrangian}
and evaluated on the background~\eqref{backgr_metric}:
\[
\mathcal{E}_A = 0, \qquad \mathcal{E}_B = 0,
\qquad \mathcal{E}_J = 0, \qquad \mathcal{E}_{\pi} = 0,
\]
where
{\small
\begin{eqnarray}
\hspace{-1cm}
&& \label{CalEa} {\cal E}_A =  F + B\pi'X'K_{X} - 2XK_{\pi} + \frac{2}{J}\left(\frac{1 - BJ'^2}{J} - (2BJ'' + J'B')\right)G_{4}
 +\frac{4BJ'}{J}\left( \frac{J'}{J } + \frac{ X'}{X} 
\right.\nonumber \\&&\left.
 + \frac{2BJ'' + J'B'}{BJ'}\right)XG_{4X} + \frac{8BJ'}{J}XX'G_{4XX} -
    B\pi'\left(\frac{4J'}{J } + \frac{ X'}{X} \right)G_{4\pi}+ 2B\pi'\left(\frac{4J'}{J } - \frac{ X'}{X} \right)XG_{4\pi X}
\nonumber\\&&
     + 4XG_{4\pi\pi}  + \frac{B\pi'}{J^2}\left(\left(1 - 3BJ'^2\right)\frac{X'}{X} - 2J'(2BJ'' + J'B')\right)
     XG_{5X}
     - \frac{2}{J^2}B^2J'^2\pi'XX'G_{5XX} 
     \nonumber\\&&
     - \frac{2}{J}\left(\frac{1 + BJ'^2}{J } + 2BJ'\frac{X'}{X} + (2BJ'' + J'B')\right)XG_{5\pi}
       - \frac{4J'}{J}B\pi'XG_{5\pi\pi} +
    4\frac{BJ'}{J}\left( \frac{J'}{J } - \frac{ X'}{X}\right)X^2G_{5\pi X}
   \nonumber\\&&
    -\frac{2B^2\pi'^3}{J^2}(5B'JJ' + B(J'^2 + 2JJ''))\pi' F_{4}
       -\frac{16 B^3 J' \pi'^3}{J} \pi'' F_4 
       + \frac{2B^3J'\pi'^5}{J} \left(B'\pi' + 2B\pi''\right)F_{4X}
\nonumber\\&&
       - \frac{4B^3J'\pi'^5}{J}F_{4\pi} , \\\nonumber
\end{eqnarray}}
{\small
\begin{eqnarray}
\nonumber
&&\label{CalEb} {\cal E}_B = F - 2XF_{X} + \left(\frac{4J'}{J } + \frac{ A'}{A} \right)B\pi'XK_{X} + 2XK_{\pi} + \frac{2}{J}\left(\frac{1 - BJ'^2}{J } - BJ'\frac{A'}{A} \right)G_{4}
\nonumber\\&&
 -\frac{4}{J}\left(\frac{1 - 2BJ'^2}{J } - 2BJ'\frac{A'}{A}\right)XG_{4X} +
    8\frac{BJ'}{J}\left( \frac{J'}{J } + \frac{ A'}{A}\right)X^2G_{4XX} - \left(\frac{4J'}{J } + \frac{ A'}{A} \right)B\pi'G_{4\pi}
\nonumber\\&&
    - 2\left(\frac{4J'}{J } + \frac{ A'}{A} \right)B\pi'XG_{4\pi X} +
    \frac{B\pi'}{J^2}\left(1 - 5BJ'^2\right)\frac{A'}{A} XG_{5X} - \frac{2B^2J'^2\pi'}{J^2}\frac{A'}{A}X^2G_{5XX}
\nonumber\\&&
     + \frac{2}{J}\left(\frac{1 - 3BJ'^2}{J } - 3BJ'\frac{A'}{A}\right)XG_{5\pi} -\frac{4BJ'}{J}\left( \frac{J'}{J } + \frac{ A'}{A}\right)X^2G_{5\pi X} -\frac{10B^3J'(A'J + AJ')\pi'^4}{AJ^2} F_{4}
     \nonumber\\&&
    +\frac{2B^4J'(A'J + AJ')\pi'^6}{AJ^2} F_{4X},\\\nonumber
\end{eqnarray}}
{\small
\begin{eqnarray}
\\\nonumber
&&\label{CalEc} {\cal E}_J = F - 2XK_{\pi} + X\left(\left(2BJ'' + J'B'\right)\frac{\pi'}{J'} + 2B\left(\pi'' - \frac{J''}{J'}\pi'\right)\right)K_{X} -
\nonumber\\&&
    \left(\frac{1}{J}\frac{\sqrt{B}}{\sqrt{A}}\left(J\frac{\sqrt{B}}{\sqrt{A}}A'\right)' + \frac{2BJ'' + J'B'}{J} \right)G_{4} - B\pi'\left(\frac{2J'}{J } + \frac{ A'}{A } + \frac{2BJ'' + J'B'}{BJ'} +
      2\frac{\pi'' - \frac{J''}{J'}\pi'}{\pi'}\right)G_{4\pi} +
\nonumber\\&&
    BX\left(-\frac{A'^2}{A^2} + \frac{2}{J}\frac{2BJ'' + J'B'}{B} + \frac{A'}{A}\frac{2BJ'' + J'B'}{BJ'} + 2\frac{A'J' + J(A'' - \frac{J''}{J'}A')}{JA}\right)G_{4X} +
\nonumber\\&&
     BX'\left(\frac{2J'}{J } + \frac{ A'}{A} \right)G_{4X} + 4XG_{4\pi \pi} +
    2B\pi'\left(\frac{2J'}{J } + \frac{ A'}{A } - \frac{ X'}{X}\right)XG_{4\pi X} + 2B\left(\frac{2J'}{J } + \frac{ A'}{A} \right)XX'G_{4XX} -
\nonumber\\&&
     \frac{1}{2}BX\left(-\frac{A'^2}{A^2} + \frac{2}{J}\frac{2BJ'' + J'B'}{B} + \frac{A'}{A}\frac{2BJ'' + J'B'}{BJ'} +
      2\frac{A'J' + J(A'' - \frac{J''}{J'}A')}{JA}\right)G_{5\pi} -
\nonumber\\&&
       BX'\left(\frac{2J'}{J } + \frac{ A'}{A} \right)G_{5\pi} - \frac{B^2J'\pi'}{2J}
     \left(2\frac{A'' - \frac{J''}{J'}A'}{A} - \frac{ A'^2}{A^2 } + \frac{A'}{A}\left(2\frac{2BJ'' + J'B'}{BJ'} + 3\frac{X'}{X}\right)\right)XG_{5X} -
\nonumber\\&&
      B\pi'X\left(\frac{2J'}{J } + \frac{ A'}{A} \right)G_{5\pi\pi} -
    \frac{BJ'}{J}\left(2\frac{X'}{X} - \frac{JA'}{A}\left( \frac{2}{J } - \frac{ X'}{XJ'}\right)\right)X^2G_{5\pi X} - \frac{B^2J'A'}{JA}\pi'XX'G_{5XX}
\nonumber\\&&
-\frac{B^2\pi'^4}{2}\left(-\frac{A'^2B}{A^2} + \frac{A'(5B'J + 2BJ')}{AJ} + 2\frac{A''BJ + 5AB'J' + 2ABJ''}{AJ}\right) F_{4}
\nonumber\\&&
 - \frac{B^3(A'J + 2AJ')}{AJ} \pi'^3 \left( 4 \pi'' F_{4} + \pi'^2 F_{4\pi} -
    \frac{\pi'^2(B'\pi' + 2B\pi'')}{2} F_{4X} \right),
\end{eqnarray}}
and
\begin{eqnarray}
&&\label{CalEphi} {\cal E}_{\pi} = \frac{1}{J^2}\sqrt{\frac{B}{A}} \cdot
\frac{\mbox{d}}{\mbox{d}r}\left[J^2 J' \sqrt{AB} \cdot
\mathcal{J}_H\right] - \mathcal{S}_H + \mathcal{JS}_{BH},
\end{eqnarray}
with
\begin{eqnarray}
&&\mathcal{J}_{H} = \frac{\pi'}{J'}F_{X} + \left( \frac{4}{J } + \frac{ A'}{AJ'}\right)XK_{X} - 2\frac{\pi'}{J'}K_{\pi} + 2\frac{\pi'}{J'}\left(\frac{1 - BJ'^2}{J^2 } - \frac{BJ'}{J}\frac{A'}{A}\right)G_{4X} -
 \nonumber \\&&
    \frac{4B\pi'}{J}\left( \frac{J'}{J } + \frac{ A'}{A}\right)XG_{4XX} - 2\left( \frac{4}{J } + \frac{ A'}{AJ'}\right)XG_{4\pi X} + \frac{1 - 3BJ'^2}{J^2}\frac{A'}{AJ'}XG_{5X}
\nonumber\\&&
    - \frac{2BJ'}{J^2}\frac{A'}{A}X^2G_{5XX} -2\frac{\pi'}{J'}\left(\frac{1 - BJ'^2}{J^2 } - \frac{BJ'}{J}\frac{A'}{A}\right)G_{5\pi} + \frac{2B\pi'}{J}\left( \frac{J'}{J } + \frac{ A'}{A}\right)XG_{5\pi X},\\\nonumber
    \\
  &&
\mathcal{S}_{H} = -F_{\pi} + 2XK_{\pi \pi} - B\pi'X'K_{\pi X} -
 \left[\frac{B}{2}\left( \frac{A'}{A} \right)^2 + \frac{2}{J}\left(\frac{1 - BJ'^2}{J } - (2BJ'' + J'B')\right)\right]G_{4\pi}
 \nonumber\\
 &&
+
      \left[\frac{B}{2}\left(2\frac{A'' - \frac{J''}{J'}A'}{A} + \frac{A'}{A}\left(\frac{4J'}{J } + \frac{2BJ'' + J'B'}{BJ'}\right)\right)\right]G_{4\pi}
\nonumber \\&&
       + B\left[\frac{X'}{X}\left(\frac{4J'}{J } + \frac{ A'}{A} \right) + \frac{4J'}{J}\left( \frac{J'}{J } + \frac{ A'}{A}\right)\right]XG_{4\pi X} +
    2\left(\frac{1 - BJ'^2}{J^2 } - \frac{BJ'}{J}\frac{A'}{A}\right)XG_{5\pi\pi}
    \nonumber\\
    &&
     - \frac{B\pi'}{J^2}\left( \frac{X'}{X } + BJ'J'\frac{A'}{A}\right)XG_{5\pi X},
\end{eqnarray}
\begin{eqnarray}
     &&
\mathcal{JS}_{BH} = -4B^2\pi'^2\left[\frac{-A'^2BJ'}{A^2J} \pi'+ \frac{J'(2A''BJ + 5AB'J' + 4ABJ'')}{AJ^2}\pi'
\right.
\nonumber\\ &&
\left.
 + \frac{A'(5B'JJ' + 3BJ'^2 + 2BJJ'')}{AJ^2} \pi' +
        \frac{6BJ'(A'J + AJ')}{AJ^2}\pi''\right]F_{4}
\nonumber\\&&
 -\frac{6B^3J'(A'J + AJ')\pi'^4}{AJ^2}F_{4\pi} +
    B^3\pi'^4\left[-\frac{A'^2BJ'}{A^2J}\pi' + \frac{J'(2A''BJ + 11AB'J' + 4ABJ'')}{AJ^2}\pi'
\right.
\nonumber\\&&
     \left.
    + \frac{A'(11B'JJ' + 3BJ'^2 + 2BJJ'')}{AJ^2}
       \pi' +  \frac{18BJ'(A'J + AJ')}{AJ^2}\pi''\right]F_{4X}
       \nonumber\\&&
 + \frac{2B^4J'(A'J + AJ')\pi'^6}{AJ^2}F_{4 \pi X} -
    \frac{B^4J'(A'J + AJ')\pi'^6(B'\pi' + 2B\pi'')}{AJ^2}F_{4XX}.
\end{eqnarray}

\section*{Appendix B}{}
In this Appendix we give the explicit expressions for coefficients
entering the quadratic action~\eqref{even_parity} for the parity
even modes.
The coefficients below involve the structures  ${\cal F}$,
${\cal G}$, ${\cal H}$, $\Xi$, $\Gamma$ and $\Sigma$, which are
the combinations introduced in
the main body of the text (see eqs.~\eqref{cal_f}--\eqref{cal_h},
~\eqref{Xi},~\eqref{Gamma} and \eqref{KSI}).
Most importantly, upon introducing beyond Horndeski term
to the Lagrangian, not only ${\cal G}$, ${\cal H}$,
$\Xi$, $\Gamma$ and $\Sigma$ get modified as it might have
been anticipated, but also additional contributions appear
in the coefficients of the action~\eqref{action_even}.
To emphasize the way beyond Horndeski terms alter the
structure of coefficients, we introduce new terms
$a_i^{BH}$ ($i=2,6,8$), $c_j^{BH}$ and $e_j^{BH}$ ($j=1,4$).
Thus, we have
\begin{eqnarray}
a_1&=&\sqrt{AB}\,\Xi,\\
a_2&=&\frac{\sqrt{AB}}{2\pi'}\left[
2\pi'\Xi'-\left(2\pi''-\frac{A'}{A}\pi'\right)\Xi
+2JJ'\left(\frac{A'}{A}-\frac{B'}{B}\right){\cal H}-4{\cal H}JJ''
\right.\\ &&\left.
+\frac{2J^2}{B}\left({\cal E}_B-{\cal E}_A\right) \right] - a_2^{BH},\\
a_3&=&-\frac{\sqrt{AB}}{2}\left(\pi'\Xi+2JJ'{\cal H}\right),\\
a_4&=&\sqrt{AB}\,{\cal H},\\
a_5&=&-\sqrt{\frac{A}{B}}J^2\frac{\partial{\cal E}_A}{\partial\pi }=a_2'-a_1'',\\
a_6&=&-\sqrt{\frac{A}{B}} \frac{1}{J\pi'} \left( J {\cal H}'+J'{\cal H}-J'{\cal F} \right) - a_6^{BH}, \\
a_7&=&a_3'+\frac{J^2}{2} \sqrt{\frac{A}{B}} {\cal E}_B,\\
a_8&=&-\frac{a_4}{2B} - a_8^{BH}, \\
a_9&=&\frac{\sqrt{A}}{J}\frac{\D}{\D r}\left(
J\sqrt{B}{\cal H}
\right)
, 
\end{eqnarray}

\begin{eqnarray}
b_1&=&\frac{1}{2}\sqrt{\frac{B}{A}}{\cal H},
\\
b_2&=&-2\sqrt{\frac{B}{A}}\Xi,
\\
b_3&=&\sqrt{\frac{B}{A}}\frac{1}{\pi'}
\left[
\left(2\pi''+\frac{B'}{B}\pi'\right)\Xi -2JJ'\left(\frac{A'}{A}-\frac{B'}{B}\right){\cal H}
+\frac{2J^2}{B}{\cal E}_A+4JJ''{\cal H}
\right]
,
\\
b_4&=&\sqrt{\frac{B}{A}}\left(\pi'\Xi+2JJ'{\cal H}\right),
\\
b_5&=&-2b_1,\\
c_1&=&-\frac{1}{\sqrt{AB}}\Xi - c_1^{BH},
\\
c_2&=&-\sqrt{AB} \left( \frac{A'}{2A} \Xi+JJ'\Gamma-\frac{J^2 \pi'}{X} \Sigma \right)
,
\\
c_3&=&J^2\sqrt{\frac{A}{B}} \frac{\partial {\cal E}_B}{\partial \pi},\\
c_4&=&\frac{1}{2}\sqrt{\frac{A}{B}} \Gamma - c_4^{BH},
\\
c_5&=&
-\frac{1}{2}\sqrt{AB}\left(\pi'\Gamma+\frac{A'}{A}{\cal H}
+\frac{2J'}{J}{\cal G}\right),
\\
c_6&=&\frac{J^2}{2} \sqrt{\frac{A}{B}} \left( \Sigma+\frac{A'B \pi'}{2J^2 A}\Xi+\frac{B\pi'J'}{J}\Gamma-\frac{1}{2} {\cal E}_B+\frac{BJ'^2}{J^2} {\cal G}+\frac{A'BJ'}{JA}{\cal H} \right),
\\
d_1&=&b_1,
\\
d_2&=&\sqrt{AB}\,\Gamma,
\\
d_3&=& \frac{\sqrt{AB}}{J^2}\left[
\frac{2JJ'}{\pi'}\left(\frac{A'}{A}-\frac{B'}{B}\right){\cal H}
-J^2\left(\frac{2J'}{J}-\frac{A'}{A}\right)\frac{\partial{\cal H}}{\partial\pi}
+\frac{2}{B\pi'}\left({\cal F}-{\cal G}\right)\right.
\nonumber\\&&\left.
-\frac{J^2}{2\pi'}\left(2\pi''+\frac{B'}{B}\pi'\right)
\left(\Gamma_1+\frac{2J'}{J}\Gamma_2\right)
-\frac{2J^2}{B\pi'}({\cal E}_A-{\cal E}_B)-\frac{4JJ''}{\pi'}{\cal H}
\right]
,
\\
d_4&=&\frac{\sqrt{AB}}{J^2}
\left({\cal G}-J^2{\cal E}_B\right),
\\
e_1&=&\frac{1}{2\sqrt{AB}}\left[
\frac{J^2}{X}({\cal E}_A-{\cal E}_B)-\frac{2}{\pi'}\Xi'+\left(\frac{A'}{A}
-\frac{X'}{X}\right)\frac{\Xi}{\pi'}
+\frac{2BJ'^2}{X}{\cal F}-\frac{2JJ'B}{X}{\cal H}' \right.\\&&\left.
-{\cal H}\frac{B^2J'^2}{JXA}\frac{\D}{\D r}\left(\frac{J^2A}{B}\right)+\frac{2BJJ''}{X}{\cal H}
\right] - e_1^{BH}, \nonumber \\
e_2&=&-\sqrt{AB}\frac{J^2}{X}\Sigma,\\
e_3&=&J^2 \sqrt{\frac{A}{B}} \frac{\partial {\cal E}_\pi}{\partial \phi},
\end{eqnarray}

\begin{eqnarray}
e_4&=&\frac{\sqrt{AB}J'^2}{8X}\left(-\frac{4{\cal G}}{J^2}-\frac{4({\cal E}_A-{\cal E}_B)}{BJ'^2}+\frac{2A'{\cal H}'}{AJ'^2}+
\frac{4{\cal G}'}{JJ'}+\frac{4}{BJ^2J'^2}\left(1-\frac{JJ'BA'}{A}\right){\cal F}
\nonumber\right.\\ &&\left.-
\frac{4{\cal H}}{BJ^2J'^2}\left(1-BJ'^2(1+\frac{2A'J}{AJ'})+J(B'J'+2BJ'')\right)-\frac{2\pi'}{J'^2}\Gamma'
\nonumber\right.\\ &&\left.
+\frac{2\pi'}{JJ'}\left(-2+\frac{A'J}{AJ'}\right)\frac{\partial{\cal H}}{\partial\phi}
+\frac{\Xi\pi'}{J^3J'}\left(2-\frac{A'J}{AJ'}\right)\left[\frac{A'BJJ'}{A}-2+2BJ'^2\right]
\nonumber\right.\\ &&\left.
+\frac{\Gamma_1\pi'}{2J'}\left[\frac{4}{J}+\frac{A'^2BJ}{A^2}-\frac{4A'}{AJ'}-\frac{4BJ'^2}{J}
-\frac{2B'}{BJ'}+\frac{2BJ''}{BJ'^2}-\frac{4\pi''}{\pi'J'}
\right]
-\nonumber\right.\\ &&\left.
-
\frac{\Gamma_2\pi'}{J'}\left[\frac{2A'}{AJ}+\frac{A'^2}{A^2J'}(1-BJ'^2)-\frac{4J'}{J^2}(1-BJ'^2)+\frac{2B'}{BJ}+\frac{4\pi''}{\pi'J}
\right]\right) - e_4^{BH},
\\
a_2^{BH}&=& 2  \sqrt{AB^3} \pi'^3 \cdot F_4 ,
\\
a_6^{BH}&=& - \sqrt{AB} \pi'^2 \left( B' \pi' + 2 B \pi''\right) \cdot F_4,
\\
a_8^{BH}&=& - \sqrt{AB^3} \pi'^4 \cdot F_4 ,
\\
c_1^{BH}&=& 4\sqrt{\frac{ B^3}{A}} J J' \pi'^3 \cdot F_4,
\\
c_4^{BH}&=& - \sqrt{\frac{ B^3}{A}} \left(A' J + 2A J'\right) \pi'^3
 \cdot F_4,
\\
e_1^{BH}&=& \frac{4}{B \pi'^2} \cdot \frac{\mbox{d}}{\mbox{d} r}
\left[\sqrt{\frac{B^5}{A}} J J' \pi'^4 \cdot F_4\right],
\\
e_4^{BH}&=& -\frac{\pi'^2}{J^2 J'} \sqrt{\frac{B^3}{A}}
\left(A'' J^2 J' - 4 A J'^3 +A' J(J'^2 - J J'')\right) \cdot F_4
\\\nonumber&&
- \frac{A'J + 2AJ'}{B J^2 J' \pi'^2} \cdot
\frac{\mbox{d}}{\mbox{d} r} \left[\sqrt{\frac{B^5}{A}} J J' \pi'^4 \cdot F_4\right].
 \end{eqnarray}

\section*{Appendix C}{}
In this Appendix we give the elements of matrices ${\cal K}_{ij}$
and ${\cal G}_{ij}$ entering the quadratic action~\eqref{even_action_final}
for parity even modes, which are used for deriving
the analytic expressions for sound speeds squared in
eq.~\eqref{speed_even}:

\begin{equation}
\begin{aligned}
  {\cal K}_{12} = {\cal K}_{21}&=
  \frac{4 A^{1/2} B^{1/2} \pi' \left({}2{\cal H} J {}J' + \Xi \pi'\right)
(\ell -1)(\ell +2) {\cal H} J'{\cal F} }
{\ell (\ell+1) A^2 J \mathcal{H}^2 \pi'^2 \left[2 J \mathcal{H} \ell (\ell+1) +\mathcal{P}_2 -4 J F_4 \ell (\ell+1)  B^2 \pi'^4\right]}
\\
& -\frac{\left. 4 A \pi' {\cal W} \left(2{\cal H} J J' + \Xi \pi'\right)^2
\left[ \ell(\ell+1) {\cal P}_1 - {\cal F} \right] \right. }
{\ell (\ell+1) A^2 J \mathcal{H}^2 \pi'^2 \left[2 J \mathcal{H} \ell (\ell+1) +\mathcal{P}_2 -4 J F_4 \ell (\ell+1)  B^2 \pi'^4\right]^2},
\end{aligned}
\end{equation}

%
%
\begin{equation}
\begin{aligned}
{\cal K}_{22}&=\frac{2 A  {\cal W}^2
\left(2{\cal H} J J' + \Xi \pi'\right)^2 \left[ \ell(\ell+1) {\cal P}_1 - {\cal F}\right]  }
{\ell (\ell+1) A^2 J^2  \mathcal{H}^2 \pi'^2 \left[2 J \mathcal{H} \ell (\ell+1) +\mathcal{P}_2 -4 J F_4 \ell (\ell+1)  B^2 \pi'^4\right]^2}
\\
&-\frac{4 A (\ell -1)(\ell +2) {\cal H} {\cal F} J'
{\cal W}
\left(2{\cal H} J J' + \Xi \pi'\right)}
{\ell (\ell+1) A^2 J^2 \mathcal{H}^2 \pi'^2 \left[2 J \mathcal{H} \ell (\ell+1) +\mathcal{P}_2 -4 J F_4 \ell (\ell+1)  B^2 \pi'^4\right]}
+ \frac{4 (\ell -1)(\ell +2) J'^2 {\cal F}}{\ell (\ell+1) A J^2 \pi'^2},
\end{aligned}
\end{equation}


\begin{equation}
\begin{aligned}
\hspace{-2cm}
\mathcal{G}_{12} = \mathcal{G}_{21 }&= -\frac{2 A^{3/2} B^{3/2} \pi' {\cal W}
\left[(\ell (\ell+1) \mathcal{P}_3  +
\mathcal{G} (\ell-1)(\ell+2) (2 \mathcal{H} J J' + \Xi \pi')^2\right]}
{\ell (\ell+1) A^2 J \mathcal{H}^2 \pi'^2 \left[2 J \mathcal{H} \ell (\ell+1) +\mathcal{P}_2 -4 J F_4 \ell (\ell+1)  B^2 \pi'^4\right]^2}\\
&+
\frac{4 B^{3/2} (\ell-1)(\ell+2) \mathcal{G} J'(2 \mathcal{H} J J' + \Xi \pi')}
{\ell (\ell+1) \mathcal{H} A^{1/2} J \pi' \left[2 J \mathcal{H} \ell (\ell+1) +\mathcal{P}_2 -4 J F_4 \ell (\ell+1)  B^2 \pi'^4\right]^2},
\end{aligned}
\end{equation}


\begin{equation}
\begin{aligned}
\hspace{-2cm}
\mathcal{G}_{22} &= \frac{A^{2} B
(\ell (\ell+1) \mathcal{W}^2 \mathcal{P}_3}
{\ell (\ell+1) A^2 J^2 \mathcal{H}^2 \pi'^2 \left[2 J \mathcal{H} \ell (\ell+1) +\mathcal{P}_2 -4 J F_4 \ell (\ell+1)  B^2 \pi'^4\right]^2} \\
&+\frac{B^{1/2}\mathcal{G} (\ell-1)(\ell+2) \left( \mathcal{W}
(2 \mathcal{H} J J' + \Xi \pi') -
2 \mathcal{H} J' [2 J \mathcal{H} \ell (\ell+1) +\mathcal{P}_2
- 4 J F_4 \ell (\ell+1)  B^2 \pi'^4]\right)^2}
{\ell (\ell+1) A^{5/2} J^2 \mathcal{H}^2 \pi'^2 \left[2 J \mathcal{H} \ell (\ell+1) +\mathcal{P}_2 -4 J F_4 \ell (\ell+1)  B^2 \pi'^4\right]^2},
\end{aligned}
\end{equation}
where
\begin{equation}
\mathcal{W} = 2 \mathcal{H} (\ell (\ell+1) + B' J J' - 2 B J'^2 + 2 B J J'')+
\Xi B' \pi'+2 \Xi B \pi'',
\end{equation}
\begin{equation}
  \mathcal{P}_3 = 2 J^2 \Gamma \mathcal{H} \Xi \pi'^2
  - \mathcal{G} \Xi^2 \pi'^2-4 J^4 \Sigma \mathcal{H}^2/B .
\end{equation}

\section*{Appendix D}{}
This Appendix contains the explicit expression for function $f_2(\pi)$,
which is found by solving eq.~\eqref{D:for_sigma}:
\[
\label{D:f2explicit}
\begin{aligned}
&f_2(\pi) = \frac{1}{8\tau^2} \:\mbox{sech}^6\left(\frac{\pi}{\tau}\right)
\left[2 w \left(- 51 + 47 \cosh\left(\frac{2\pi}{\tau}\right)\right)
\cdot \cosh^2\left(\frac{\pi}{\tau}\right) 
\cdot \mbox{sech}\left[ u + \sinh\left(\frac{\pi}{\tau}\right)\right]
\right.\\&\left. 
+ 8 w \cdot \cosh^4\left(\frac{\pi}{\tau}\right) 
\cdot \mbox{sech}\left[\sinh\left(\frac{\pi}{\tau}\right)\right]
\cdot \mbox{sech}^2\left[ u + \sinh\left(\frac{\pi}{\tau}\right)\right] 
\cdot \sinh\left(\frac{\pi}{\tau}\right)
\Big(6 \sinh(u) 
\right.\\&\left.
+ \sinh\left[ u + 2 \sinh\left(\frac{\pi}{\tau}\right) \right]\Big)
+ w \cdot \mbox{sech}\left[ u + \sinh\left(\frac{\pi}{\tau}\right)\right] 
\Bigg(9 \sinh^2\left(\frac{2\pi}{\tau}\right) 
\right.\\&\left.
+ 2 w\cdot \mbox{sech}^2\left[\sinh\left(\frac{\pi}{\tau}\right)\right] 
\cdot \mbox{sech}^3\left[ u + \sinh\left(\frac{\pi}{\tau}\right)\right] 
\cdot \sinh(u) \cdot
\Bigg(4\cdot \sinh(u) 
+ 2 \sinh\left(u-\frac{2 \pi}{\tau}\right)
\right.\\&\left.
- 2 \sinh\left(u-\frac{\pi}{\tau}\right) 
+ 2 \sinh\left(u+\frac{\pi}{\tau}\right) 
+ 2 \sinh\left(u+\frac{2 \pi}{\tau}\right)
+ 2 \sinh\left[u+2 \sinh\left(\frac{\pi}{\tau}\right)\right]
\right.\\&\left.
+ \sinh\left[u- \frac{2 \pi}{\tau} + 2 \sinh\left(\frac{\pi}{\tau}\right)\right]
- 2 \sinh\left[u- \frac{\pi}{\tau} + 2 \sinh\left(\frac{\pi}{\tau}\right)\right]
+ 2 \sinh\left[u + \frac{\pi}{\tau} + 2 \sinh\left(\frac{\pi}{\tau}\right)\right]
\right.\\&\left.
+ \sinh\left[u+ \frac{2 \pi}{\tau} + 2 \sinh\left(\frac{\pi}{\tau}\right)\right]\Bigg)\Bigg)
+ \cosh^6\left(\frac{\pi}{\tau}\right) \Big(-11 + 
16 \sinh^2\left(\frac{\pi}{\tau}\right) + 3 \sinh\left(\frac{\pi}{\tau}\right) 
\right.\\&\left.
\times \mbox{tanh}\left[\tau \sinh\left(\frac{\pi}{\tau}\right)\right]\Big)
\right]. 
\end{aligned}
\]


\end{document}